\documentclass[pra,twocolumn,amsmath,amssymb]{revtex4-2}

\usepackage{amsmath}

\usepackage[caption=false]{subfig}
\usepackage{bbm}
\usepackage[utf8x]{inputenc}

\usepackage{mathtools}
\usepackage{bbold}
\usepackage{subcaption} 
\usepackage{appendix}
\usepackage{xcolor}
\usepackage{float}
\usepackage{placeins}
\usepackage{braket}
\usepackage{slashed}
\usepackage{graphicx}
\definecolor{darkred}{rgb}{0.4,0.0,0.0}
\definecolor{darkgreen}{rgb}{0.0,0.3,0.0}
\definecolor{darkblue}{rgb}{0.0,0.0,0.7}
\usepackage[bookmarks,linktocpage,colorlinks,
linkcolor = darkred,
urlcolor  = darkgreen,
citecolor = darkblue]{hyperref}
\def\beq{\begin{equation}}
  \def\enq{\end{equation}}
\usepackage{lineno}
\usepackage{comment}
\definecolor{winered}{rgb}{0.8,0,0}
\definecolor{darkb}{rgb}{0,0,0.8}

\usepackage[english]{babel}

\begin{document}

\title{Finite size scaling of bitstring probability distributions for Rydberg arrays} 
\author{Zane Ozzello$^{1}$}
\author{Avi Kaufman $^{1}$}
\author{Yannick Meurice$^{1}$}
\affiliation{$^1$ Department of Physics and Astronomy, The University of Iowa, Iowa City, IA 52242 USA }
\def\lt{\lambda ^t}
\def\note{note}
\def\beq{\begin{equation}}
  \def\enq{\end{equation}}

\date{\today}

\begin{abstract}
We calculate the probabilities  $p_{\{n\}}$ of  the measured bitstrings $\{n\}$ for the vacuum of Rydberg ladders with $N_q$ atoms. As $N_q$ increases, the $p_{\{n\}}$ decrease but become more dense in the low $p$ region, raising the possibility that their smallness could be compensated by their large number.
The importance of the low probability states can be estimated from the cumulative probability distribution $\Sigma(p_{\Lambda},N_q)$, which is the probability to observe any state having a probability $p\leq p_{\Lambda}$. For not too large values of $p_{\Lambda}$, it is possible to approximately collapse the $\Sigma(p_{\Lambda},N_q)$ for successive $N_q$ into a function that resembles the Fermi function when plotted as a function of $-\ln(p_{\Lambda})$. We show that the number of shots necessary to reduce $\Sigma(p_{\Lambda},N_q)$ to a sufficiently low value grows exponentially with $N_q$. We discuss the implications for calculating observables associated with the vacuum.  We show how the accuracy of these observables may or may not be robust against the number of shots.  The collapsed probabilities occur in other systems, where they typically are on one side of the phase transition, in a basis dependent way.  
   
\end{abstract}

\maketitle

\def\pla{p_{\Lambda}}
\def\nr{N_{r}}
\def\npc{\mathcal{N}(p)}
\def\dpc{\frac{\Delta p}{2}}
\def\nsh{N_{sh}}

\section{Introduction}
In quantum simulation \cite{Altman_qsimarch, Bauer_qsim4hep, Boixo_qsupreme, Arute_qsupreme}, it is necessary to collect a large number of measurements to estimate observables with decent statistical errors.  These simulations are broadly relevant across particle, nuclear, and condensed matter physics, where quantum simulation provides a direct avenue for the observation of quantum behavior \cite{Xu_glueball, Balaji_su3, Huffman_u1, Brokemeier_multimaj,Costa_nucanneal, Singh_nucres,  Kirmani_hall, Calajo_qmbscar, Klco_schwinger, Balantekin_qineut}.  The large number of measurements becomes increasingly more important as the size of the simulated systems increases.  These larger simulations open more possibilities for error, so knowing the minimal number of shots by which to obtain an intended statistical accuracy is invaluable.  It has previously been shown that by characterizing a system with large scale distributions, such as with the bitstring states and probabilities comprising the system, conjectures can be drawn about the overall structure and behavior of the system \cite{Shaw_bdpt, Guhr_rmtrev, Mark_thermalhse, Mok_randombits, Kaufman_diagnose}.  

In this work, we take large sampling shot totals from matrix product states (MPS), calculated with density matrix renormalization group (DMRG) methods, providing large amounts of bitstring ($\{n\}$) probability ($p_{\{n\}}$) dictionaries.  These dictionaries were previously validated in \cite{Kaufman_diagnose} and \cite{Kaufman_filter}.  This study is primarily conducted with arrays of Rydberg atoms, qubit systems with ground ($\ket{g}$) and Rydberg ($\ket{r}$) states. These are rich systems well documented in analog quantum computing, with a range of interest across many areas of research \cite{Surace_lgtstring,Notarnicola_1dqed, Celi_2dqgt, Meurice_abhiggs, Senseman_qupyth, Fromholz_triangle, Gonzalez-Cuadra_2dstringbreak, Halimeh_coldatomqsim, Gonzalez-Cuadra_nagt_qudit, Choi_quantchaos, Mark_plasmahep, Fresnelteam_solventpredict, Kaufman_filter, Turner_rydscar, Huang_qptu1, Mukherjee_rydperiod, Ebadi_256, Allen_simulate, Ebadi_independentset, Wang_directiontransport}.  We are able to quantify the behavior of systems based on these aforementioned dictionaries by investigating various probabilistic properties as the system size $N_q$, the number of qubits, increases.  We know that with an increase in $N_q$ a $p_{\{n\}}$ will decrease in value and increase in density in lower probability regions.  However, we want to quantify the ways in which behaviors like this scale with $N_q$ explicitly.  

Without any extra analysis, looking at the maximal probabilities, $p_{max}$, reveals an exponential scaling with the system size.  This is a natural limitation, as when the number of shots is less than $1/p_{max}$ then any given state can only occur at most 0 or 1 times in the distribution.  The dependence on shot total is the result of the large uncertainty in the estimation of individual probabilities.  We then refine this analysis, using the cumulative probability distribution, $\Sigma(p_\Lambda, N_q)$, which quantifies the probability of observing a state with probability $p\leq p_\Lambda$.  The inflection point of this quantity also displays exponential scaling with the system size.  Given a sufficiently small $p_\Lambda$ the $\Sigma$ curves can be collapsed onto a single curve of the Fermi function form, assumed to be plotted as a function of $-ln(p_\Lambda)$.  The inflection points previously discussed fall near to where the probabilities add to $1/2$, i.e., $\Sigma(p_\Lambda)=1/2$.  This allows the parent Fermi form to be used to easily argue for scaling across system sizes at an intermediate region of the cumulative probability distribution curves.  

This article is organized as follows.  In Section II, we highlight our main system of investigation, arrays of Rydberg atoms, and then introduce cumulative probabilities and show the collapse of their curves.  In Section III, we divide the cumulative probability distributions into four regions.  In Region I we quantify how a system's maximal probability scales with system size, in Region II we ascribe the Fermi form to the curve and quantify the inflection point's scaling, in Region III we look at a different form of the collapse of the cumulative probability distributions and compare these fits to those of the Fermi form, and in Region IV we quantify how single-occurrence states scale with system size.  In Section IV,  we discuss how the reliance of different observables on shot total is related to how reliant the individual observable is on probability distributions.  We see there is a distinct difference between the accuracy of estimates of the occupation and the classical mutual information to experimental shot total, as the former is more dependent on the dominant states whereas the latter is impacted by the entire distribution's makeup.  In Section V, we show that cumulative distributions apply to other models and highlight that the Fermi form is most applicable only within certain parameter ranges of any given model.  In the Appendices we provide additional detail to fitting arguments,demonstrate the applicability of our methods to other parameter ranges of Rydberg systems in brief, and demonstrate how cumulative distributions may be able to inform optimal shot totals with the goal of approximating entanglement with bitstring results.

\section{Collapse of $\Sigma$} \label{sec:collapse}

The application of cumulative probabilities distributions is demonstrated here with arrays of Rydberg atoms.  Such a system is described by the Hamiltonian
\begin{align}
    \label{eq:ryd_ham}
    \hat{H} &= \frac{\Omega}{2}\sum_i(\ket{g_i}\bra{r_i} + \ket{r_i}\bra{g_i}) \\ \notag
    &-\Delta\sum_i  \hat{n}_i +\sum_{i<j}V_{ij}\hat{n}_i \hat{n}_j.
\end{align}
Here, $\Delta$ is the detuning frequency, $\Omega$ is the Rabi frequency, $V_{ij}$ is a van der Waals interaction potential, $\hat{n}_i=\ket{r_i}\bra{r_i}$, and $\ket{g} \text{ and } \ket{r}$ are the ground and Rydberg states, respectively.  Then $V_{ij}=(\Omega R_b^6)/|\vec{r}_i-\vec{r}_j|^6$, where $\vec{r}_k$ are the positions of the Rydberg atoms and $R_b$ is the Rydberg blockade radius, where inside of which two atoms are disfavored from being energetically excited which occurs when $V=\Omega$.  In order to work with the array in a ladder configuration, we use an array of two rows of atoms where the vertical spacing of the atoms is twice that of the horizontal spacing, giving us a ratio between the horizontal and vertical spacing of $\rho=2$.  We take the lattice spacing to be $a=1$, and unless otherwise stated, the results are shown at $R_b/a=2.00$ and $\Delta/\Omega=3.5$.  The Rydberg ladder system and its states have well-studied properties and a rich phase structure \cite{Zhang_crit, Zhang_floating}, making it a natural choice to first investigate cumulative probability distributions and particularly to look at these parameter ranges.  For all Rydberg systems shown in this work, the data come from $10^9$ samplings of an MPS representing each system, mimicking large shot totals directly from quantum device simulation.   

A simple way to quantitatively characterize the collective importance of low probabilities is to calculate the cumulative probability distribution. For the problems considered here, we use the notation $\{n\}$  for the bitstrings and $p_{\{n\}}$
for their associated probability. We define the cumulative probability as a function of the maximal probability $\pla$ as 
\begin{equation}
\label{eq:sigma}
    \Sigma (\pla)=\sum_{\{n\}: p_{\{n\}}\leq \pla }p_{\{n\}}\ , 
\end{equation}
which represents the probability of observing any state with a probability that is at most $\pla$.
In the following, we study the cumulative probability distribution for the bitstrings corresponding to the ground state of ladders of Rydberg atoms. We use the convention where the $j$-th atom, $n_j$ = 0 for the ground state and 1 for the Rydberg state. 

\begin{figure}
    \centering
    \includegraphics[width=8.6cm]{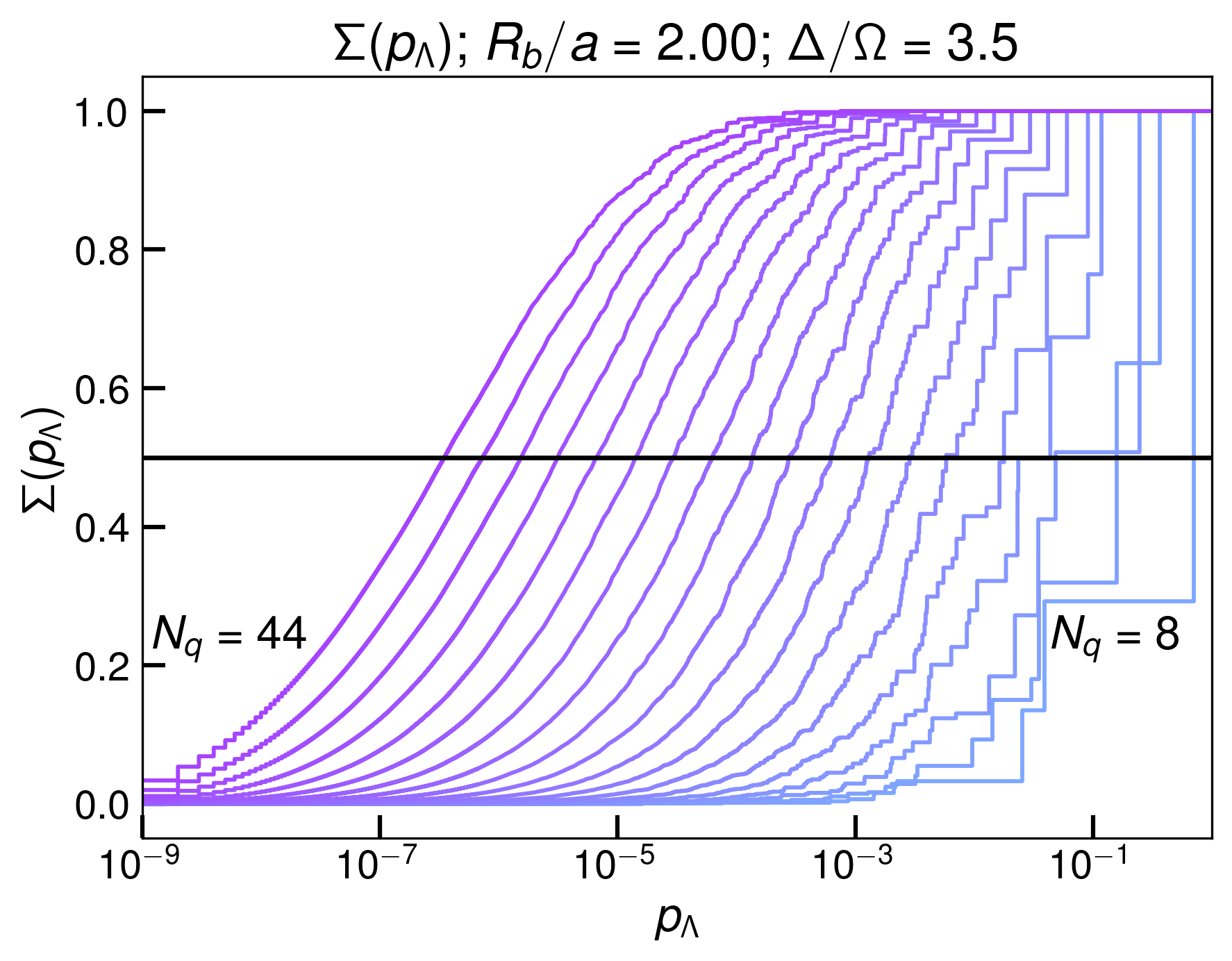}
    \caption{Cumulative distributions across different Rydberg ladder systems of sizes 8 qubits (4 rungs) to 44 qubits (22 rungs).  }
    \label{fig:cd_linlog}
\end{figure}

A cumulative probability gives the probability of observing a state with a probability less than or equal to $p_\Lambda$.  Using a quantity like this immediately contextualizes the state content for a system --- it informs if, for a given parameter or system size, one should expect that the description is dominated by a handful of high likelihood states with a large amount of low likelihood or vise versa.  This is displayed in Figure \ref{fig:cd_linlog}.    

The upshot comes from looking at the cumulative probability distributions across system sizes.  There is apparent repeated behavior across increasing system sizes, given a sufficiently large system size of around $N_q\geq20$, especially in the $0.2<\Sigma(p_\Lambda)<0.8$ range.  In this region, across all the system sizes, apart from the smallest, the distributions look approximately equivalent to each other.  It is worth noting that if a large cumulative probability is considered, there is not this smooth behavior across systems, but this more describes how the dominant states for the system behave and at what probability they occur.  Whereas at low cumulative probability, that behavior better gives an idea of the low occurrence states and their diversity.   

\begin{figure}[h]
    \centering
    \includegraphics[width=8.6cm]{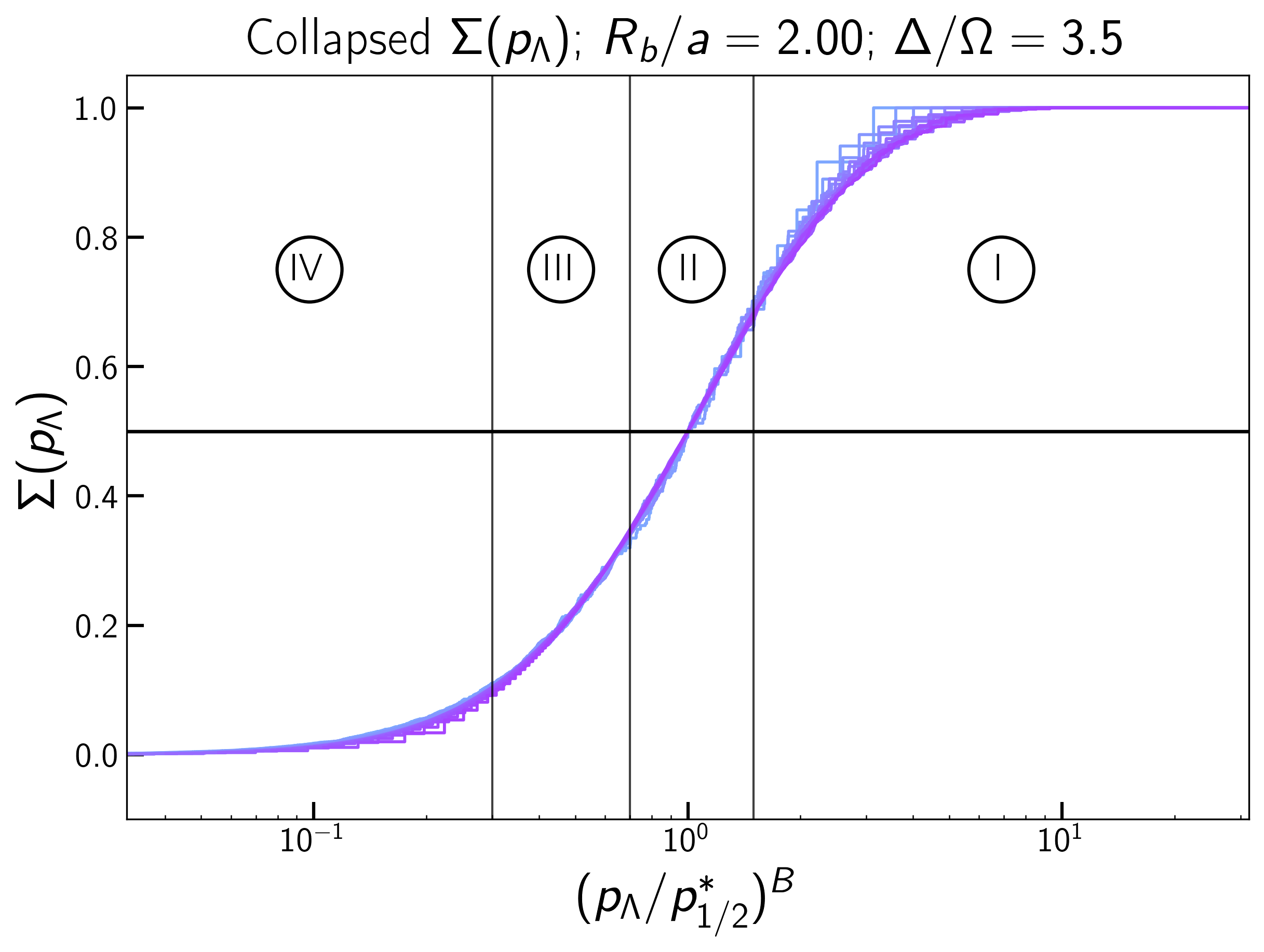}
    \caption{Collapsed cumulative distributions with four regions shown.  }
    \label{fig:linlog_collapse}
\end{figure}

When looking specifically at the cumulative distributions of Rydberg ladders, with $R_b/a=2.00$, of sizes in the range of 20 to 44 atoms, it is observationally evident that there is shared behavior among all distributions regardless of system size. Specifically, these distributions have near linear behavior in their middle.  By choosing a central $\Sigma(p_\Lambda)$ value, these distributions can then be fitted with a function of a linear form (when on a linear-log plot).  This repeated behavior is demonstrated in Figure \ref{fig:cd_linlog} and applying the linear behavior for collapse is shown in Figure \ref{fig:linlog_collapse}.  Further discussion will be held in the Region II subsection.

Given that the distributions are able to be fit neatly with the linear function, this behavior can be used to collapse all of these lines onto one another, essentially collapsing all the distributions into a similar form.  For each distribution, there is a $p_\Lambda$ value at which the distribution reaches the fixed $\Sigma(p_\Lambda)$ value.  In this work, we will investigate in Region II, $\Sigma(p_\Lambda)=1/2$, and in Region III, $\Sigma(p_\Lambda)=0.01$, in more detail.  The value of $p_\Lambda$ where a specific value of $\Sigma$ is reached will be referred to as $p^*$.  Specifically, if $\Sigma=\alpha$, then the $p^*$ where that happens will be referred to as $p^*_\alpha$.  The distributions of different system sizes will collapse by taking $p_\Lambda$ to $(p_\Lambda/p^*)^B$ where $B$ is the slope of the line fit to each individual distribution and then plot the distributions against this adjusted independent variable.  In Figure \ref{fig:linlog_collapse} this is shown expressly.  Note that the central region fits near exactly, whereas both the formerly high and low $p_\Lambda$ regions do not fall as neatly.  Regardless, this highlights the striking similarity across system sizes.  

This is where the power of the cumulative probability distributions is most evident.  While not necessarily perfect, all distributions (outside of small systems) collapse onto a single line and could all be roughly described by the same distribution.  Therefore, a cumulative probability from the collapsed distribution can describe the state behavior of any system.  This means that the behavior of a large system will be predictable from that of smaller systems.

\section{Regions of $\Sigma$}
Looking at Figure \ref{fig:cd_linlog}, there are four regions of different behavior that hold across different system sizes.  There is a high $\Sigma$ region that has discrete steps due to the dominant probabilities, there is then a smooth linear region that is used for the fitting, there is then a smaller, but still smooth, region that can be fitted well on a logarithmically scaled plot, and then there is the tail occupied by low-occurrence states.  

These regions are specifically highlighted for the collapsed distribution curves in Figure \ref{fig:linlog_collapse} to give an idea of how they are split.  In the collapse, you better see how different regions have different behavior when plotted, as these areas on the collapse plot do not fall as neatly on the single collapsed curve.  Depending on the overall system size, the exact bounds on region size may fluctuate.  For example, Region I will be larger for smaller systems, but will have more room for Region III.  However, a larger system will give a smaller Region I, but also provide a smaller Region III.  In the coming subsections, we will highlight information gleaned by looking at each of these Regions on their own, and discuss how quantities from the regions inform behavior as system size increases. 

All subsequent analysis of the different regions is done with $R_b/a=2.00$.  It is generally applicable to other parameter regimes, and a condensed look at similar results is shown in Appendix \ref{app:300} for $R_b/a=3.00$.  

\subsection{Region I}
\begin{figure}
    \centering
    \includegraphics[width=8.6cm]{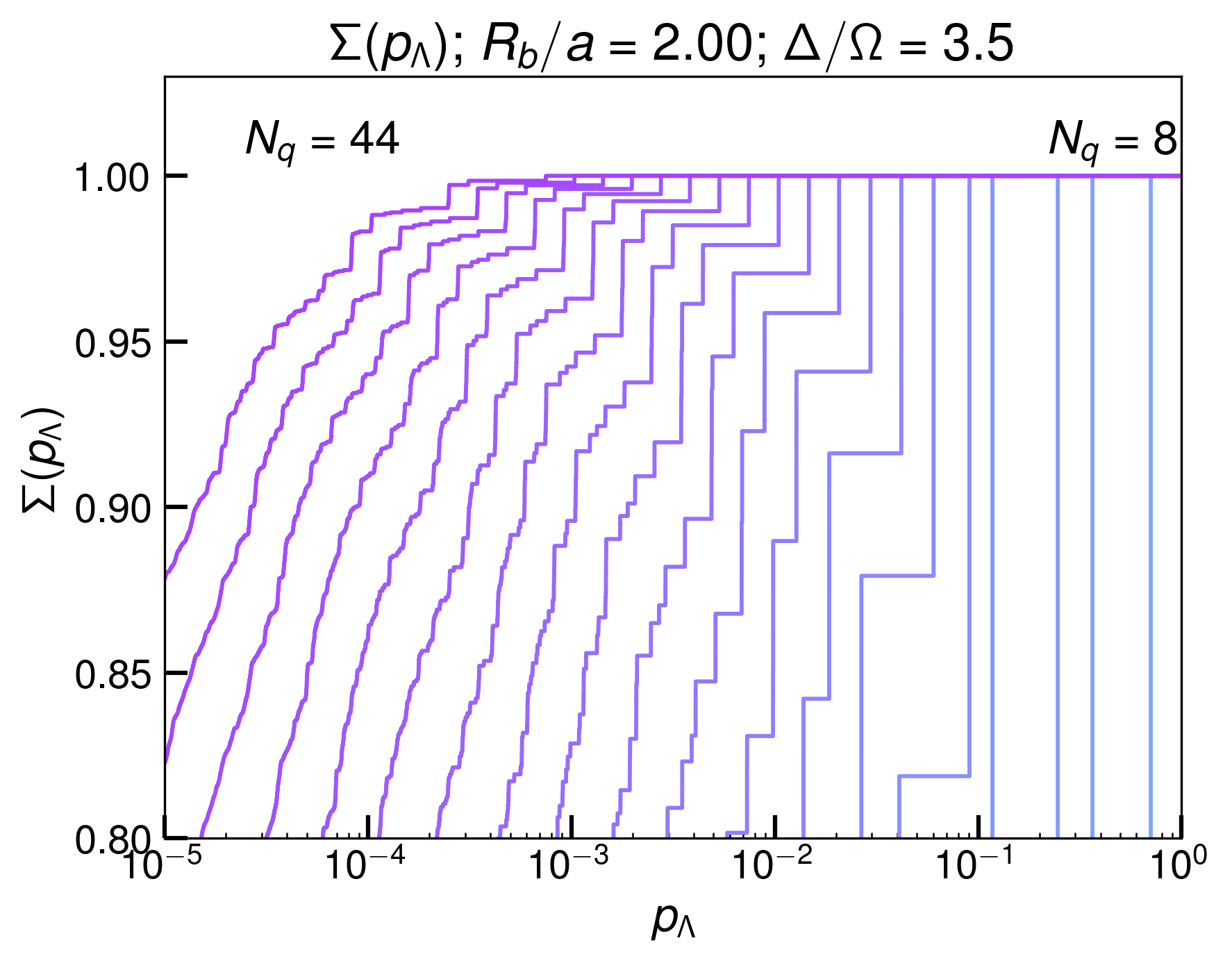}
    \caption{Cumulative distributions across different system sizes, zoomed in view at larger $p_\Lambda$ values}
    \label{fig:cd_reg1_linlog}
\end{figure}

We begin by looking at Region I (Figure \ref{fig:cd_reg1_linlog}), a region that visually has large steps in its cumulative distribution.  These large jumps are due to potentially degenerate large probability states that dominate behavior at that system size.  Such steps make it difficult for this Region to be fitted directly like later Regions; however, this does not prevent regularities across system sizes.  Despite system size, each Rydberg array configuration has a dominant state which contributes to the bulk of the cumulative probability.   With system size, this dominant state has its $p_\Lambda$ decrease and contributes less to the cumulative probability as system size increases.  

\begin{figure}[t]
    \centering
    \includegraphics[width=8.6cm]{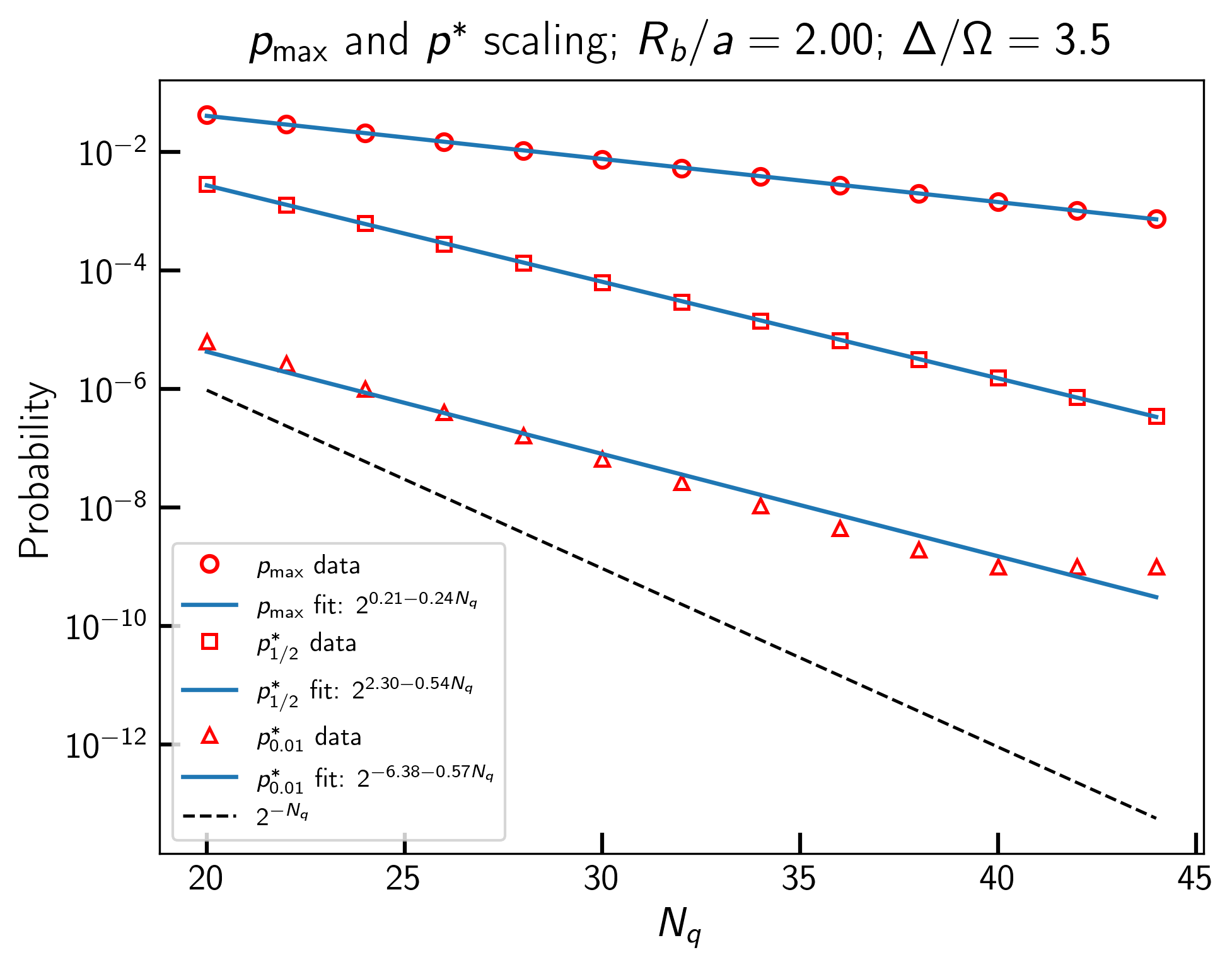}
    \caption{$p_{max}$ and $p^*$ scaling compared against scaling with system size}
    \label{fig:scaling}
\end{figure}

In Figure \ref{fig:scaling}, these dominant states are tracked exactly.  We will refer to such states as $p_{max}$, the largest single probability state at a given system size.  This quantity clearly fits to a base-two exponential on a logarithmic y-scale plot.  With this, our qualitative observation from before is confirmed:  with increasing system size, the maximal $p_{\{n\}}$ decreases in value.  This exposes a level of limitation corresponding to the number of overall shots of an experiment.  Given $N_{sh}$ shots taken for an experiment, the minimal probability that could be supported is $1/N_{sh}$.  According to Figure \ref{fig:scaling}, there would exist a $p_{max}\sim 1/N_{sh}$, which would functionally be the limit of system size that could be investigated at such a shot total where there is a sufficiently large $N_{sh}$ such that the $p_{n}$ could be faithfully reconstructed.  For our discussion here, we are working with one billion sampling shots taken from an MPS, so based on the predicted fit, the maximal system size that could be simulated would approach 120 qubits, which still matches the limits of accessible quantum devices.  

\subsection{Region II}\label{subsec:reg2}
In Region II, the cumulative distribution becomes much smoother and more well-behaved.  This is the area utilized to get the curves from Figure \ref{fig:cd_linlog} to collapse onto the shared distribution in Figure \ref{fig:linlog_collapse}.  This is done by taking advantage of the linear behavior in this region to then rescale and shift all curves upon one another.  Generally, we are able to fit the entire distributions with the form (given a $p^*$ where the fit is centered):
\begin{equation}
    \label{eq:full_fit_sig}
    \Sigma(p_\Lambda)\simeq \frac{(p_\Lambda/p^*)^\beta}{(1+(p_\Lambda/p^*)^\beta)}.
\end{equation}

Taking $-ln(p_\Lambda) \rightarrow \epsilon$ and $\rightarrow -ln(p^*) \rightarrow \mu$ the cumulative probability distributions can be well-approximated by Fermi-Dirac(FD) statistical distributions of the form:  
\begin{equation}
    \label{eq:fermi}
    \Sigma\simeq\frac{1}{1+e^{\beta(\epsilon-\mu)}}.
\end{equation}
These cumulative probability distributions are not perfectly described by Fermi-Dirac statistics; however, the model does provide a means for discussing the behavior of the distributions, especially in Region II.  This is used in Figure \ref{fig:linlog_collapse}, where the $B$ used in the collapse is the FD $\beta$.  

The FD nature of the cumulative distribution is also used in Figure \ref{fig:scaling}.  In FD statistics there exists an $\epsilon$ such that the occupation is half filled.  Equivalently, such a quantity exists for the cumulative probability distributions; there is a $p_\Lambda$, which will be referred to as $p^*_{1/2}$, such that $\Sigma(p^*_{1/2})=1/2$.  This gives an idea of how the probability distribution is filling up.  At small system sizes $p^*_{1/2}$ is a higher value, indicating that the total probability is dominated by high probability states, as supported by the $p_{max}$ scaling also in Fig. \ref{fig:scaling}, and it takes more of the total states in the system to halfway fill the system.  As system size increases, $p^*_{1/2}$ decreases, again indicating there is a uniform way in which the probability distributions of the systems fill.  As before, making the y-scale logarithmic in Figure \ref{fig:scaling} gives a linear fit to the data.  Keeping the shot total for an experiment in mind, the scaling again reveals a limit to the system size with faithful reconstruction that is achievable with a specific shot total.  Here, using the modeled fit it is expected for $p^*_{1/2} \sim 1/N_{sh}$ the system size where the largest probabilities are faithfully resolved would be at a system size of around 60 qubits.

\subsection{Region III}

\begin{figure}
    \centering
    \includegraphics[width=8.6cm]{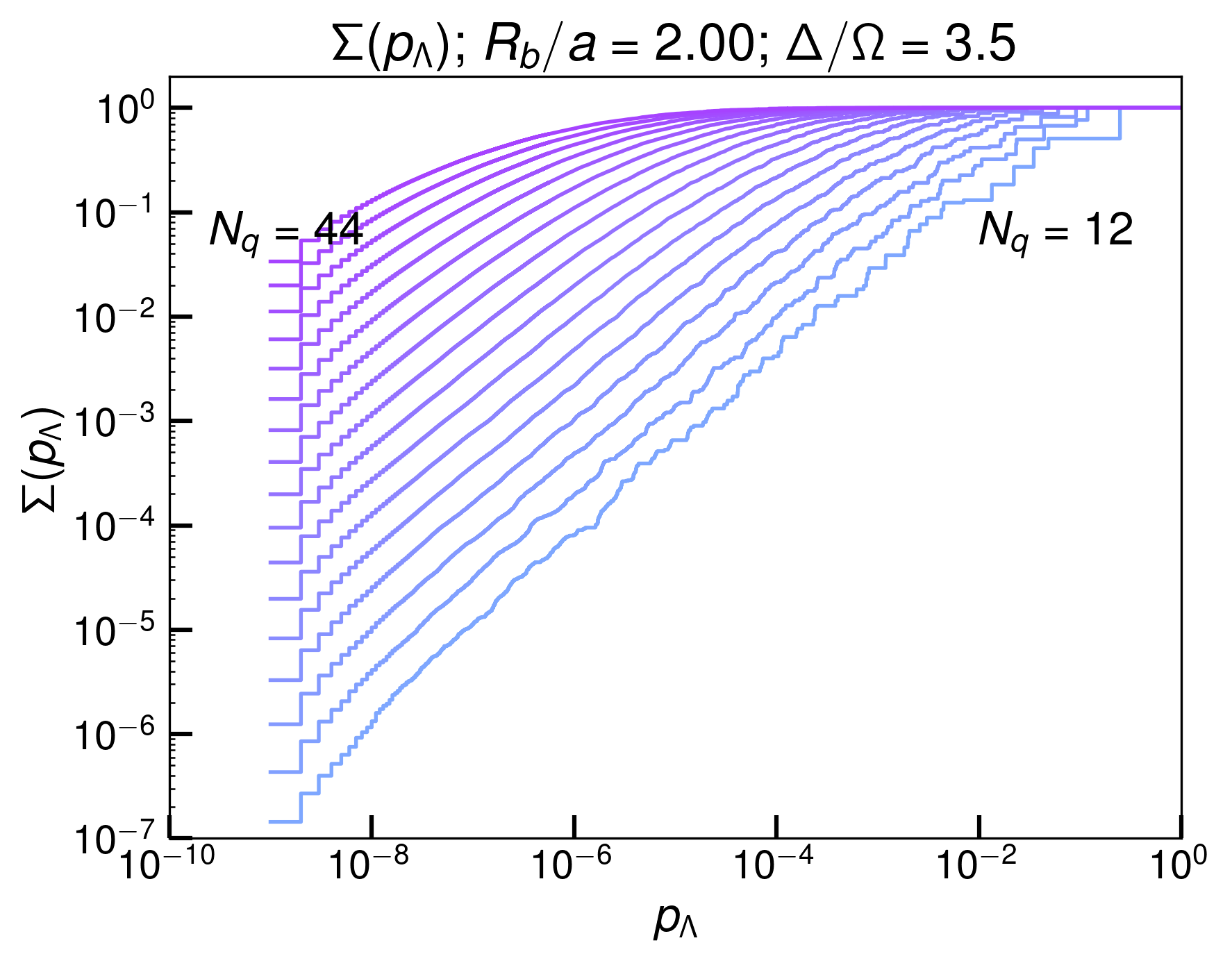}
    \caption{Region III, log-log plotted cumulative distributions.  }
    \label{fig:cd_loglog}
\end{figure}

\begin{figure}
    \centering
    \includegraphics[width=8.6cm]{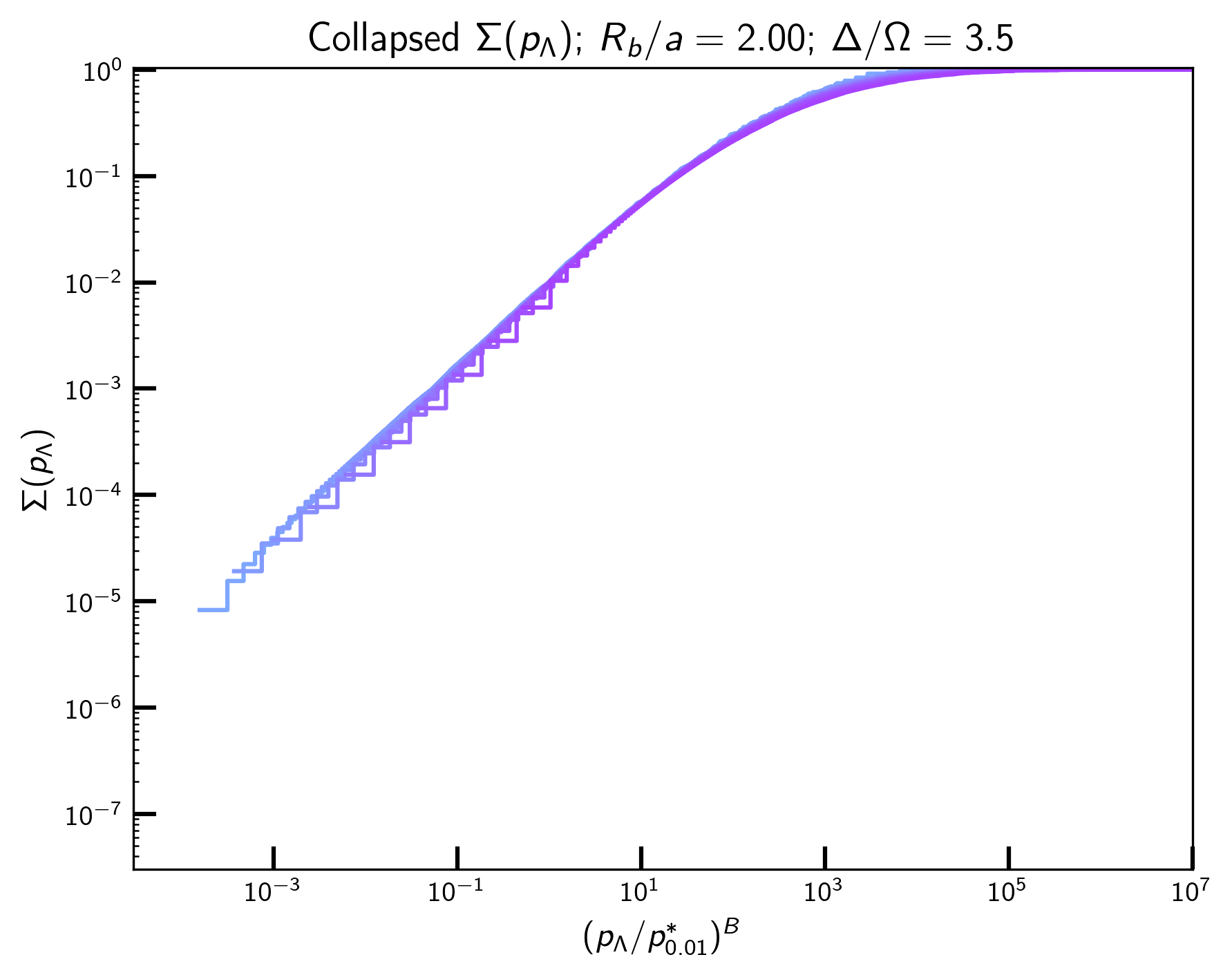}
    \caption{Collapse based on Region III; $p^*_{0.01}$.}
    \label{fig:loglog_collapse}
\end{figure}

When looking at Region III, the cumulative probability distributions will need to be plotted with logarithmic scales on the $x$ and $y$ axes to best reveal the finer details of the curves.  This Region is the low $\Sigma$ values that exist before the single occurrence states.  Looking at Figure \ref{fig:cd_loglog}, this Region seems to have some uniform behavior across system size, not unlike what was seen in Region II, taking advantage of this, a collapse is attempted as was done in \ref{sec:collapse}, yielding Figure \ref{fig:loglog_collapse}.  Taking inspiration from the previous discussion on the FD distribution, we refer to the points we fit around as $\Sigma(p^*_{y})=y$ where $y$ is the accumulated probability to that $p_\Lambda$.  In Region III, the cumulative probability distribution is fit around $p^*_y$, now using the form:
\begin{equation}
    \label{eq:reg3fit}
    \Sigma(p_\Lambda)\simeq A p_\Lambda^B.  
\end{equation}

While the collapse is clean, the rescaling of the distributions is not as forgiving.  The larger system size distributions are dramatically rescaled to align on a single curve.  This is unsurprising given how the different cumulative distributions fall with respect to each other in Figure \ref{fig:cd_loglog}.  Seemingly, the curves stack upon each other with an apparent even spacing, as opposed to moving more to the left and sloping over one another like they appear to in Figure \ref{fig:cd_linlog}.  This is due to the logarithmic y-scale exposing much more of the inherent low probability behavior to Region III.  This separation is directly a result of the finite shot total.  Recall that all of this data is collected from  $10^9$ sampling shots of an MPS with DMRG.  It is known that due to this shot total there cannot be a probability smaller than $10^{-9}$, which is why all of the cumulative distributions abruptly stop at this point.  Were there to be more shots, eventually the true state would be reconstructed where the distributions extend to much smaller $p_\Lambda$ while sloping down as before.  

Region II and Region III are comparable in terms of behavior, and this is evidenced through a comparison of their fit functions in Equations (\ref{eq:full_fit_sig}) \& (\ref{eq:reg3fit}).  Given this, we take 
\begin{equation}
    \label{eq:reg2vreg3}
    (p_\Lambda/p^*)^\beta = A p_\Lambda^B.  
\end{equation}
Where the comparison is obvious, $B\simeq\beta$ and $A\simeq(p^*)^{-\beta}$.  There is a further look at these slope parameters and a look at how fitting choices impact their value in both Regions in Appendix \ref{app:slope}.  

Given that a $p^*$ can be found for this fit function, we can also observe its scaling with system size, just as in Region II.  Here, we use $p^*_{0.01}$ and are able to compare to our other scalings in Figure \ref{fig:scaling}.  This lower $p^*$ value scales more quickly than our previous larger value, $p^*_{1/2}$.  This makes sense given that the cumulative probabilities in Region III grow ever closer to the behavior of Region IV as system size increases, implying that the limits of finite $N_{sh}$ becomes more prevalent.  This will be explained further in the discussion of Region IV.   

The limited nature of Region III, which is dependent upon shot total, provides another place to inform experiment.  The $\Sigma$ value in Region III can be treated as an error estimate.  Assume that there exists a cumulative distribution for the state to be simulated.  Then a $\Sigma=z$ can be selected where $z$ is the desired error; using Figure \ref{fig:cd_loglog} as a reference, choose a $z$.  Then the $p_\Lambda$ at which the cumulative probability distribution crosses this value of $\Sigma$ indicates the minimal amount of shots that would be required to have this accuracy.  For example, using Figure \ref{fig:cd_loglog} for a 20 qubit Rydberg ladder system with a desired error of $10^{-2}$ informs that about a minimum of $10^5$ shots are required given $p^{min}_\Lambda=1/N_{sh}$.

\subsection{Region IV}
\label{subsec:reg4}

In Region IV, the cumulative probability distributions are strongly affected by the imperfect resolution of probabilities $10^{-9}$ and lower.  This is directly a result of the chosen sampling shot totals; increase the shots and Region III will be extended.  However, with a finite shot total, Region IV will always exist.  Were a state to be perfectly reconstructed with a suitable amount of shots there would still be discrete singular steps for the lowest probability states.  However, eventually all cumulative probability distribution curves will trend towards and reach zero, unlike how they can only approach it in our plots.  The finite shot totals prevent perfect reconstruction, so there will be abrupt cutoffs in the curves upon reaching a $p_\Lambda$ limit from $N_{sh}$.   

\begin{figure}
    \centering
    \includegraphics[width=8.6cm]{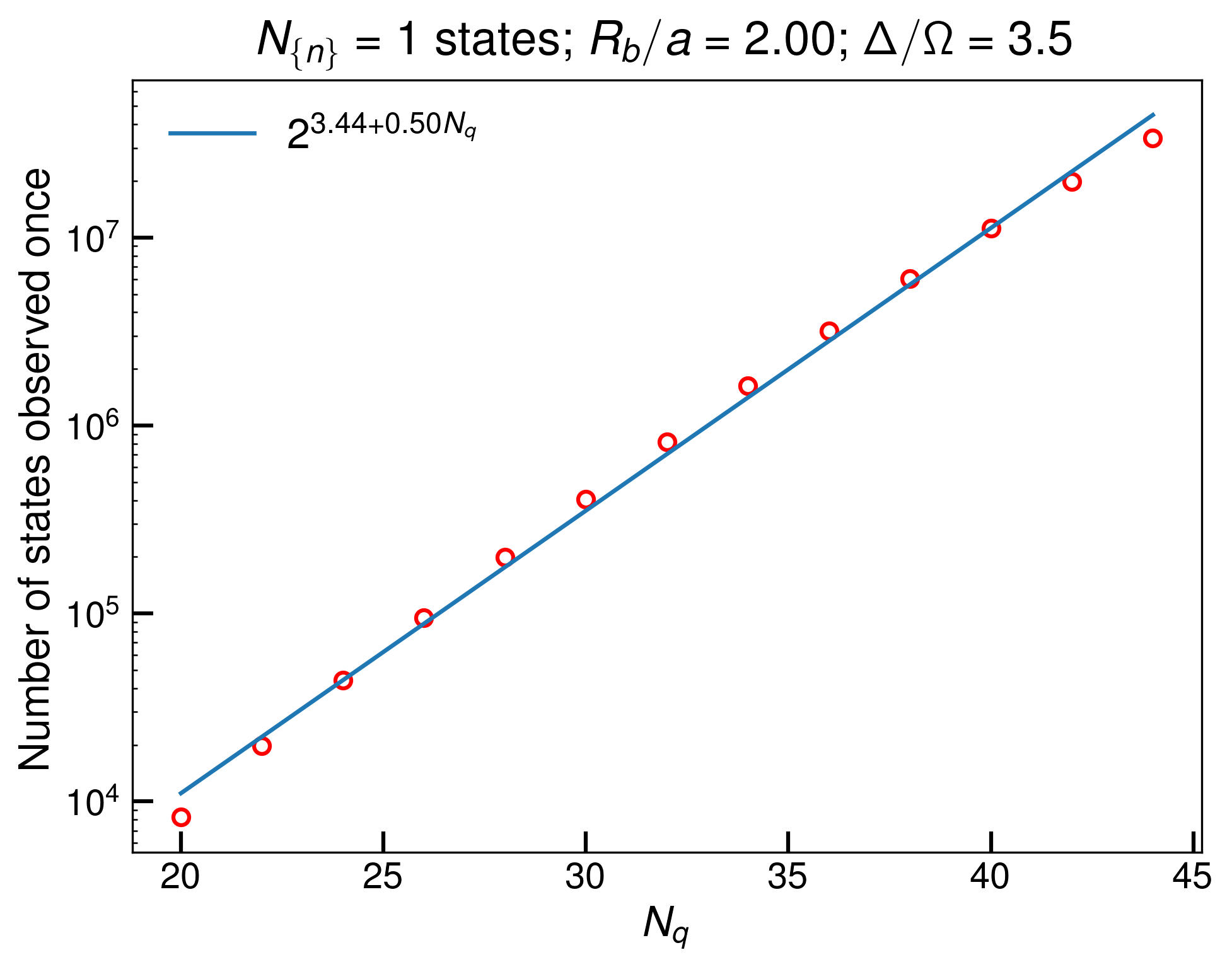}
    \caption{Single occurrence states with system size}
    \label{fig:singleton}
\end{figure}

Despite this, we can still use the information from Region IV to help estimate the ignorance in what states are missing from the reconstruction.  This is done in Figure \ref{fig:singleton}, which shows the number of states that occur exactly once in the $10^9$ sampling shots compared against system size.

This provides an approximation to the integrated probabilities which the cumulative distributions are not accessing i.e.
\begin{equation}
\label{eq:singleton_approx}
\Sigma(10^{-9})\simeq\frac{\text{\# single occurrence states}}{N_{sh}=10^{9}}.
\end{equation}
This number of single occurrence states does increase with system size, but does grow slower than $2^{N_q}$.  In fact, the number of states is not linear with system size and it should plateau when the number of states observed once approaches $10^9$, which still provides room for this shot total to be effective up to 60 qubits.  

\section{Impact on Observables}
It is natural to then discuss the impact that shot total has on observable values.  Observables are the goal result from quantum simulation, so it is beneficial to have a means of making informed shot total choices.   We will investigate this in two settings:  one where the observable quantity is dependent on the entire probability distribution for a state where small changes in single probabilities greatly impact the overall quantity, and another where the observable is dependent on a few dominant outcomes so the probabilities are robust against minor differences.  To consider the effect of shot total on observables, we will perform a resampling procedure where we take a selected amount of shots from the parent dictionary of bitstrings and their accompanied shot totals such that a faithful distribution of that shot total is constructed.  This allows for testing in what will amount to trials with lower shot totals.  We then average over multiple resampling trials to better display what occurs at a specific resampling value. 

\begin{figure}
    \centering
    \includegraphics[width=8.6cm]{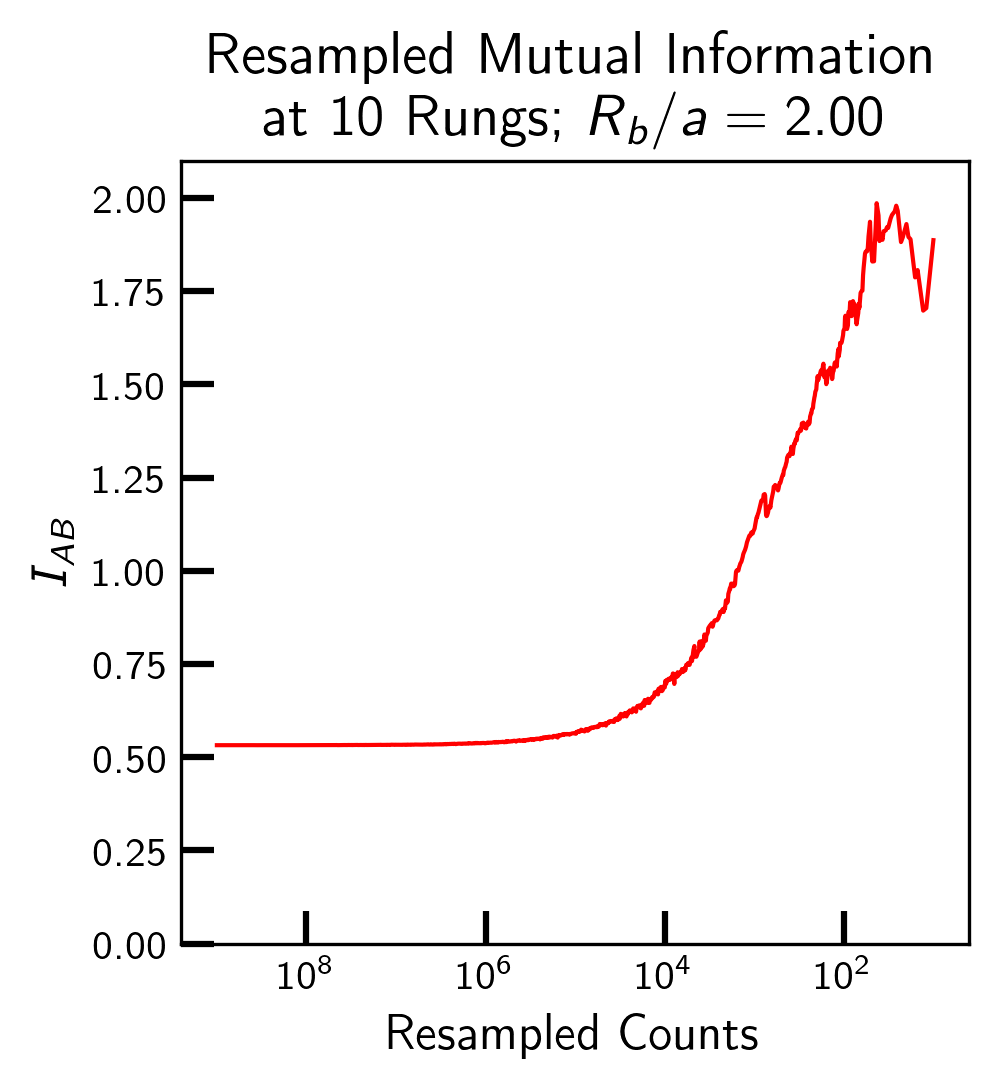}
    \caption{Mutual information with number of resamples for ten rungs.}
    \label{fig:mi}
\end{figure}

It has been previously discussed \cite{Kaufman_diagnose,Kaufman_filter, Meurice_cmi} that the classical mutual information can be used as an approximation to the von Neumann entanglement entropy and it was shown that its value is responsive to the probability distribution it is calculated over.  The classical mutual information is defined as
\begin{equation}
\label{eq:mi}
I_{AB}\equiv S_A^X+S_B^X-S_{AB}^X,
\end{equation}
where the Shannon entropy over marginal probabilities, $p_{\{n\}_A}=\sum_{\{n\}_B}p_{\{n\}_A \{n\}_B}$ where $\{n\}_R$ denotes the bitstring describing region $R$ in a partition, is

\begin{equation}
\label{eq:margin_shannon}
S^X_{A}\equiv -\sum_{\{n\}_A} p_{\{n\}_A}\ln(p_{\{n\}_A}).
\end{equation}
So, to calculate a classical mutual information value, we evaluate Eq. \ref{eq:mi} for a given bitstring probability distribution.  To reiterate, given that the entire distribution is used in calculation, any differences for individual states can change the resultant value.  For ten rungs, and resampling down from $10^9$ counts, this will look like Fig. \ref{fig:mi}.  

To discuss the sampling error on observables that do not depend on the fine details of the probabilities in the distribution, we consider an observable $\hat O$ which is diagonal in the computational basis,

\begin{equation}
\langle \{n\} | \hat O | \{n'\} \rangle
=
O_{\{n\}}\,
\delta_{\{n\}\{n'\}} .
\end{equation}

\begin{figure}
    \centering
    \includegraphics[width=8.6cm]{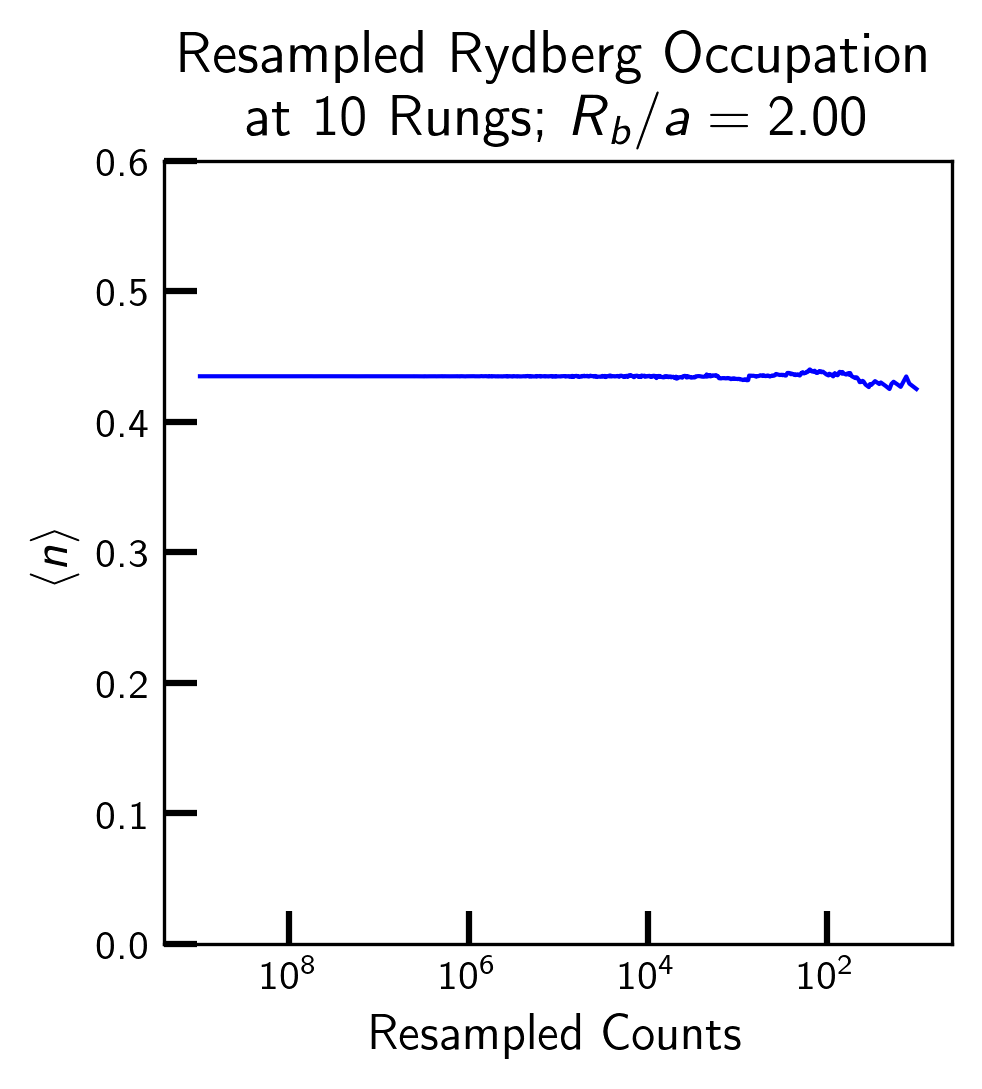}
    \caption{Rydberg occupation at ten rungs.  }
    \label{fig:occ}
\end{figure}

The expectation value in the vacuum state is

\begin{equation}
\langle \mathrm{vac} | \hat O | \mathrm{vac} \rangle
=
\bar O ,
\end{equation}

with

\begin{equation}
\bar O
=
\sum_{\{n\}}
p_{\{n\}}\, O_{\{n\}} .
\end{equation}
We take the simple example of the average Rydberg occupation,  $\hat{O}=\langle n\rangle=\frac{1}{N_q}\sum_{j=1}^{N_q}n_j$, which takes only $N_q$ values.  This quantity can have multiple states that describe the same occupation, as there are multiple configurations to get a given amount of atoms with Rydberg occupation.  For observables such as this, changes in some states' probabilities may not affect the overall value, as these different states may mask discrepancies.  Using this set up, we can look at the observable values with probabilities that are drawn from the resampled dictionaries, just as was done for the classical mutual information.  This is shown in Figure \ref{fig:occ}.

Comparing Figures \ref{fig:mi} and \ref{fig:occ}, the classical mutual information has dramatic responses to the size of the resampled dictionary, whereas the Rydberg occupation is relatively robust.  This matches expectation, given that the classical mutual information is reliant on the overall probability distribution it would naturally be more responsive to changes in that distribution.  However, it is worth noting that both quantities show the most response, relatively, to a resampled probability distribution size at small amounts of resampling counts, which is also to be expected for the small sizes, which are more likely to have errors in capturing the probability of multiple individual states.  

\begin{figure}
    \centering
    \includegraphics[width=8.6cm]{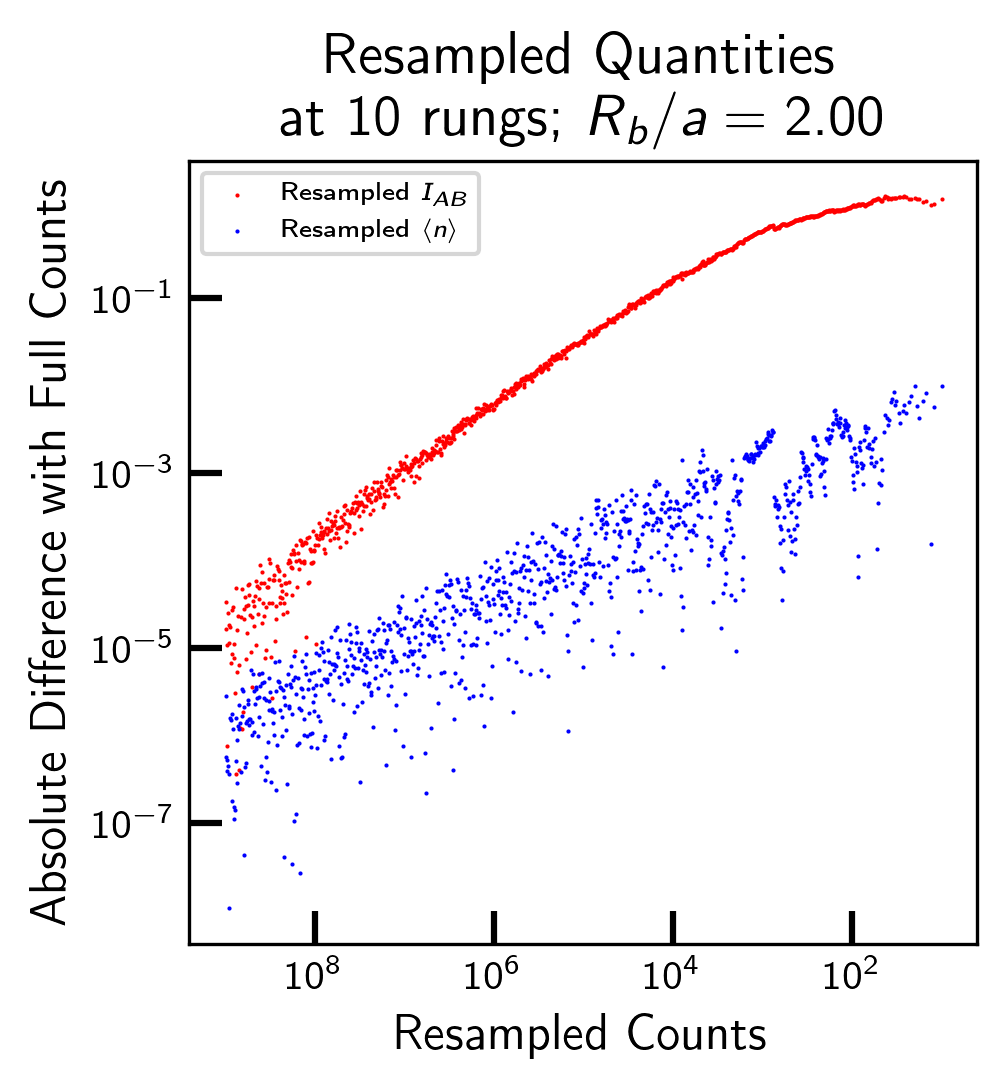}
    \caption{Absolute difference for the observables at the resampled count size and the full counts value ($10^9$).}
    \label{fig:absdiff_all}
\end{figure}

To get a better idea of how these observable quantities differ from the large probability distribution values, we look at the absolute difference between the full and resampled distributions.  The absolute difference of observable values from their full count ($10^9$) value is shown in Figure \ref{fig:absdiff_all}.  In the figure, there is clearly larger dispersal in resampled occupation values as opposed to those of the resampled classical mutual information.  To explain this, we will take $10^3$ different resampled dictionaries at $10^6$ counts and recalculate the observables for each resampling to get an idea of how these observables may change based on the resampled distribution.
\FloatBarrier
\begin{figure}
    \centering
    \includegraphics[width=8.6cm]{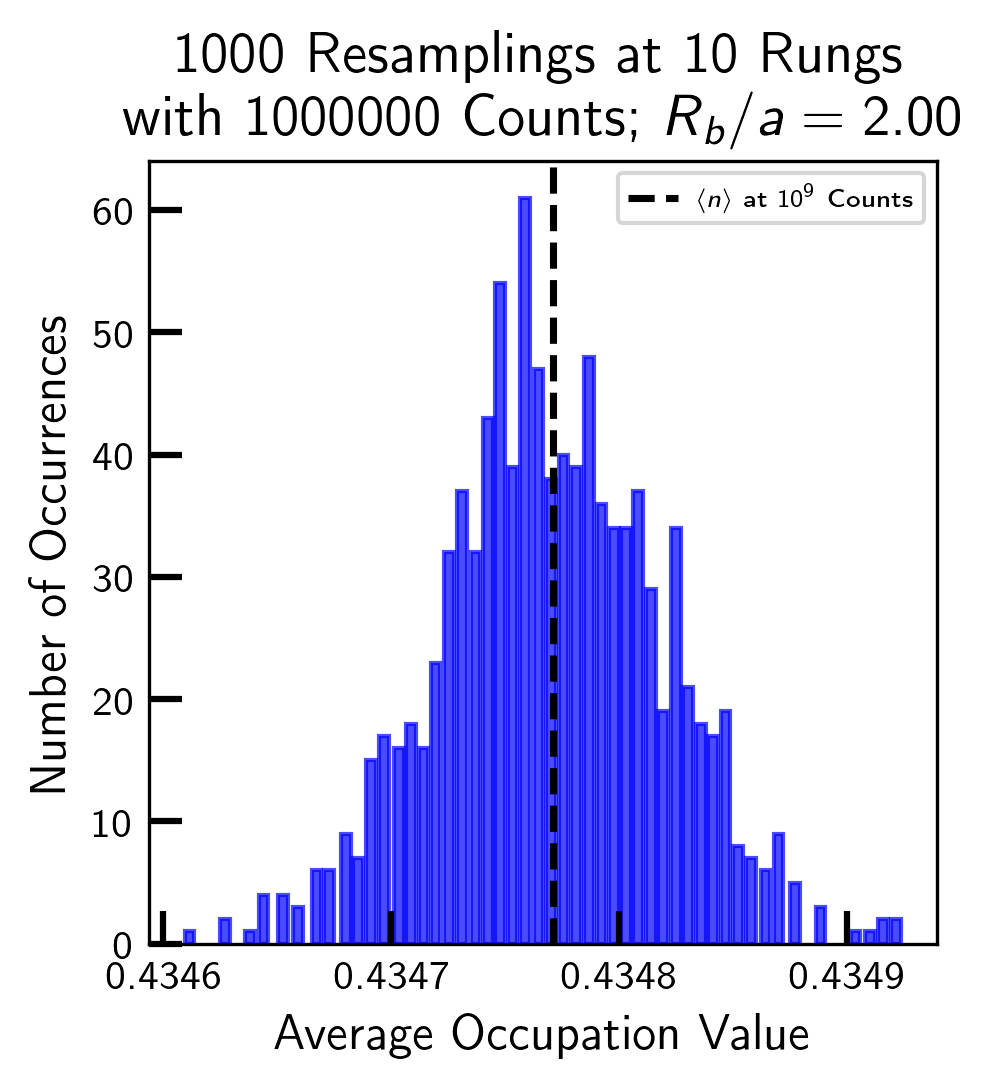}
    \caption{1000 resamplings of 1000000 counts for the Rydberg occupation.  }
    \label{fig:occ_resample}
\end{figure}

\begin{figure}
    \centering
    \includegraphics[width=8.6cm]{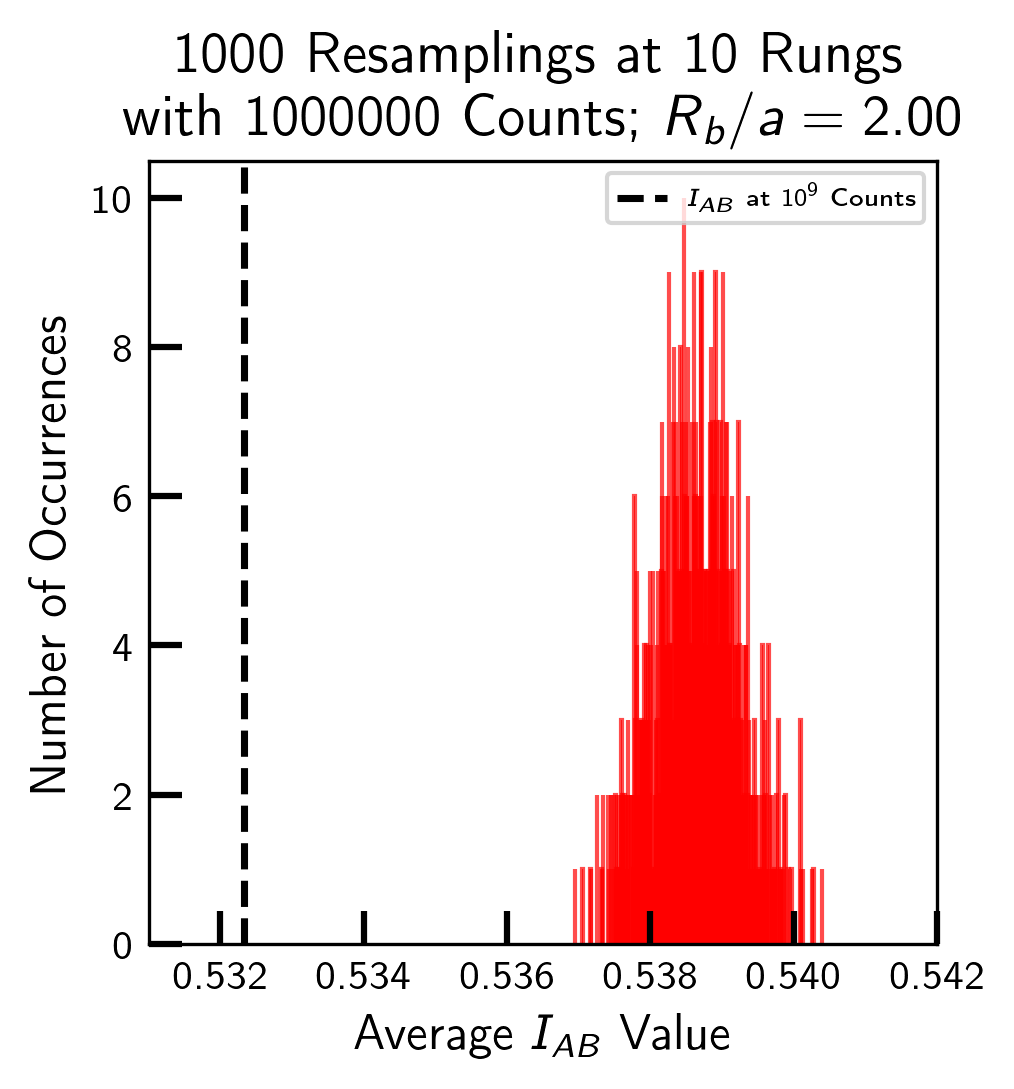}
    \caption{1000 resamplings of 1000000 counts for the classical mutual information.  }
    \label{fig:mi_resample}
\end{figure}

For the Rydberg occupation, this is shown in Figure \ref{fig:occ_resample}.  Here, the various resamplings form a distribution centered around the full counts value.  However, for the classical mutual information in Figure \ref{fig:mi_resample} a similar distribution forms, but the full count value does not center the distribution, but rather occurs at a lower value separate from the distribution.  This is behavior not dissimilar to the behavior of the filtering procedure \cite{Kaufman_filter}, where not using all possible states affects the classical mutual information value. However, resampling provides a different effect than that of the filtering process.  While resampling does cause an increase in classical mutual information value with smaller total counts, it increases more rapidly, experiences more drops, and is due to inexact reconstructions of the exact state as opposed to just casting out the lower probability states.   
The wider spread in Figure \ref{fig:absdiff_all} for the Rydbeg occupation comes from the location of the full count value and the distance it lies from the distribution.  Recall that Figure \ref{fig:absdiff_all} shows the absolute difference from the full count value and the resampled value.  Figure \ref{fig:occ_resample} shows that the difference would be expected to be somewhere around the full count value according to the distribution, but Figure \ref{fig:mi_resample} indicates that the expected classical mutual information value will have a near fixed distance from the full count value.  These effects are especially exaggerated on the log-log scaling of Figure \ref{fig:absdiff_all}.   

 Figure \ref{fig:absdiff_all} also provides insight into the error on the different observables.  For both the classical mutual information and the Rydberg occupation, the absolute difference decreases as the number of counts increases.  This matches what is known and expected:  there are more accurate results for larger shot totals.  The figure also shows that the occupation is much more resilient to shot total than the classical mutual information.  The Rydberg occupation has approximately lower absolute error across all resampling counts and seems to show that the absolute difference with the full counts value goes like $~\sqrt{N_{sh}}$.  This matches the expectation discussed before, that observables more dependent on the individual bitstring probabilities, like the classical mutual information, will show more error with differing count totals.

 \section{Other Models}
The usage of cumulative probability distributions and the model fitting of significant probability quantities applies in other models beyond Rydberg systems. Here, we highlight cumulative distribution results for the transverse field Ising model (TFIM) in 1+1D for system sizes of 12 to 21 sites, with a more modest $10^6$ shots, also retrieved from sampling an MPS that was found using DMRG.

The Hamiltonian we simulate is: 
\begin{equation}
    \label{eq:ising_zz}
    \hat{H}_{zz} = - h \sum_{j=1}^{N_s} \hat{\sigma}^x_j -J \sum_{j=1}^{N_s} \hat{\sigma}^z_{j} \hat{\sigma}^z_{j+1}.  
\end{equation}
This Hamiltonian can also have its basis changed which then gives 
\begin{equation}
    \label{eq:ising_xx}
    \hat{H}_{xx} = - h \sum_{j=1}^{N_s} \hat{\sigma}^z_j -J \sum_{j=1}^{N_s} \hat{\sigma}^x_{j} \hat{\sigma}^x_{j+1}.  
\end{equation}

\begin{figure*}
    \centering
    \includegraphics[width=0.48\linewidth]{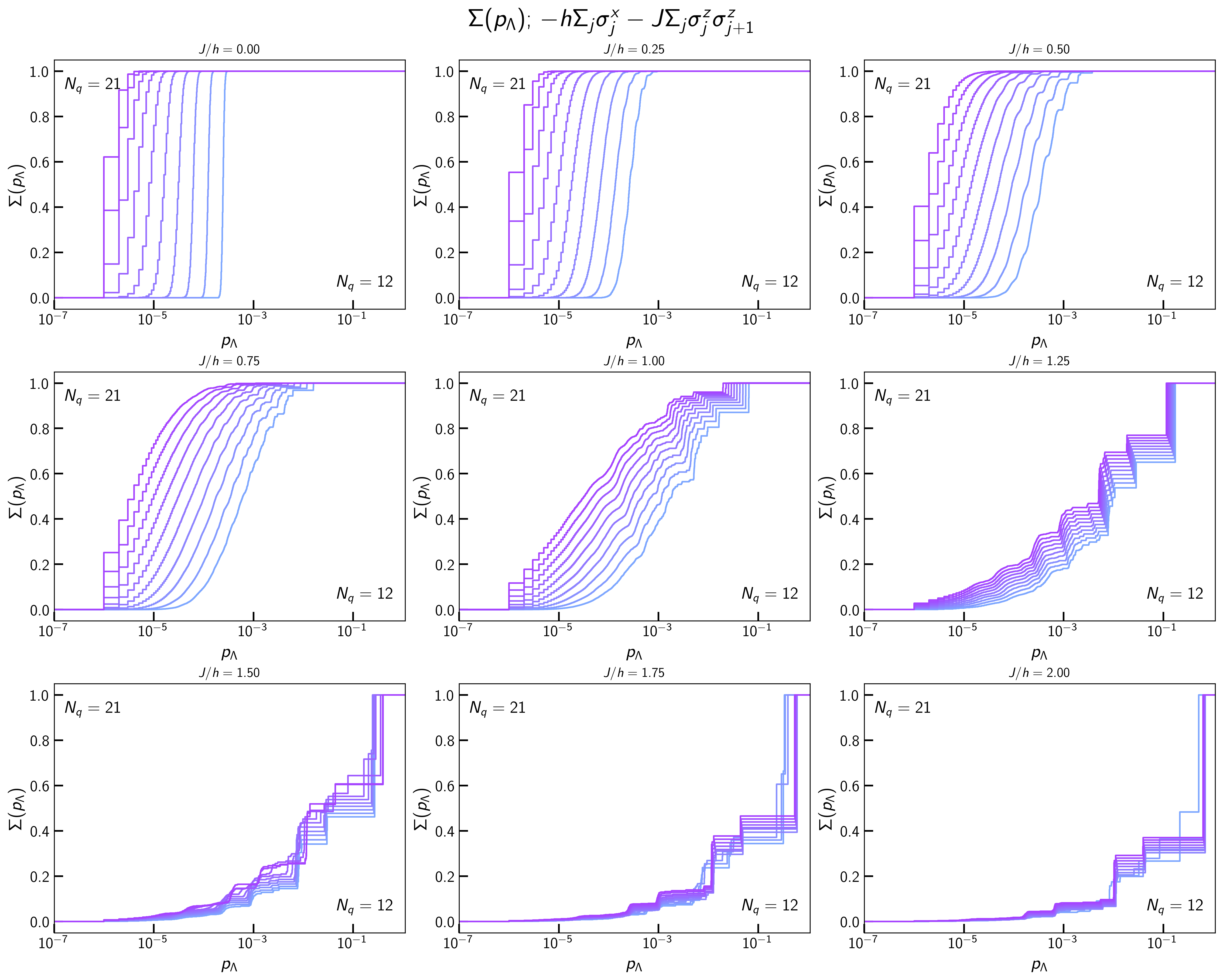}
    \hfill
    \includegraphics[width=0.48\linewidth]{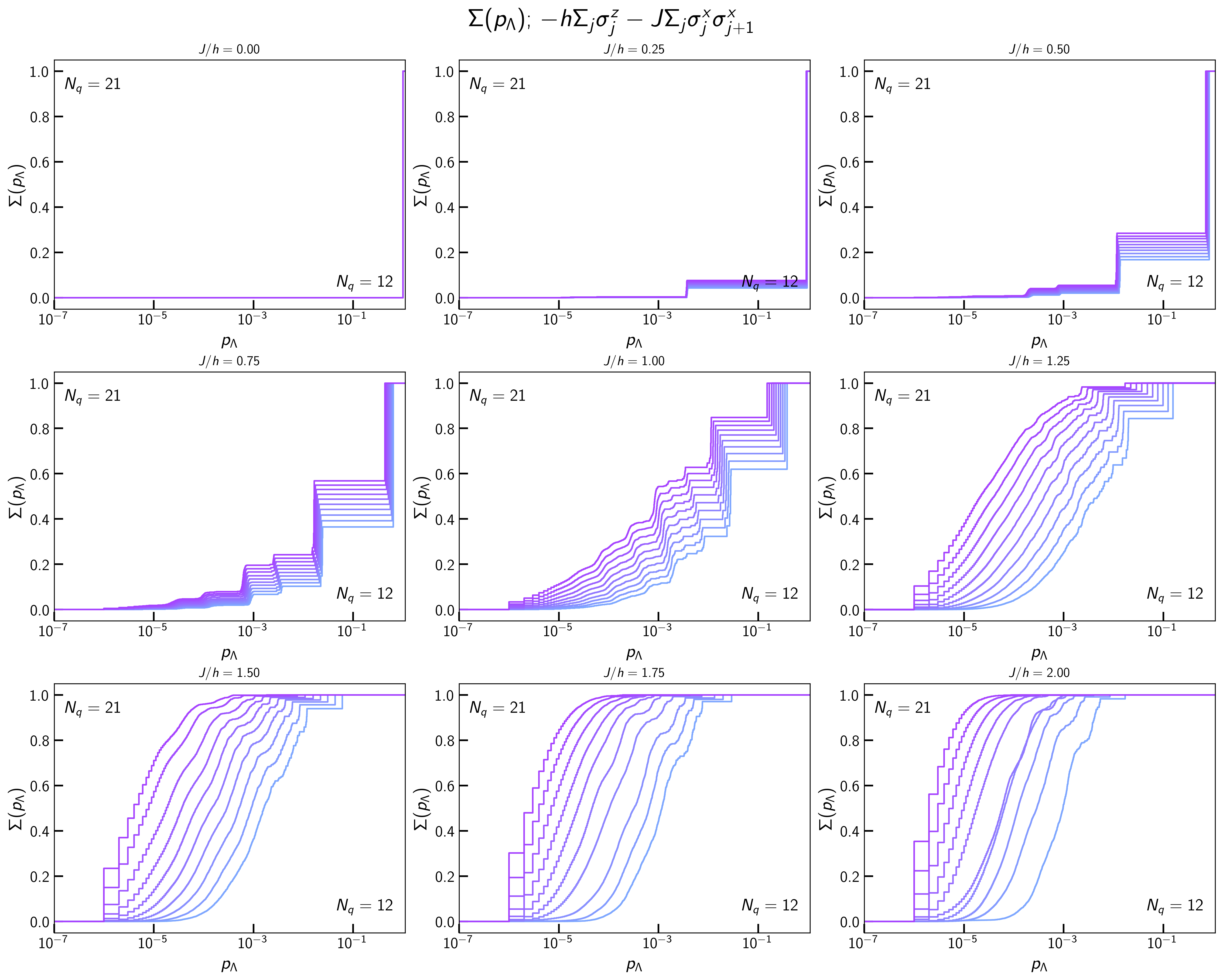}
    \caption{Left: Cumulative distributions from 12 to 21 sites for $10^6$ shots for the $H_{zz}$ Hamiltonian. Right: The same, but for $H_{xx}$ Hamiltonian.}
    \label{fig:ising}
\end{figure*}

The TFIM has a well-known phase transition at $J/h=1$.  When in the disordered phase, low $J/h$, $H_{xx}$ is the more natural basis to operate in, whereas in the ordered phase, high $J/h$, that basis is the $H_{zz}$.  This amounts to choosing the basis in which the dominant term is diagonal in the phase. We see how this extends to our study of cumulative probability distributions in Figure \ref{fig:ising}.  

The cumulative distributions for either basis display an inverse behavior.  For the vacuum at $J/h=0$ in $H_{xx}$ every shot yields the same state independent of system size; this comes from the ground state being either all $\ket{+z}$ or all $\ket{-z}$, but never a superposition, giving a ground state of all spins up or down in the $z$-direction.  As the coupling increases, slowly more states are possible and different system sizes differentiate, until on the other side of the transition we see distributions across system sizes reminiscent of $R_b/a=2.00$ where we were able to apply the FD form.  Continuing to increase the coupling towards an infinite limit will cause the curves to stay separate, but have jumps at $2^{-N}$, as each spin site can either be up or down.  The coupling is not pushed this high in the figure, however looking at the cumulative distributions for increased coupling in the alternate basis we see that it will still come true.

In the distributions for $H_{zz}$, at low coupling each curve is trending towards a single step at $2^{-N}$ as detailed before.  In this different basis, there is inherently a superposition where the states are $\ket{\pm x}=\frac{1}{\sqrt{2}}(\ket{+z}\pm\ket{-z})$.  The ground state will still pick to be $\ket{+x}$ or $\ket{-x}$, but since they are a superposition, for a system of size $N_q$, the cumulative probability distribution will jump at $2^{-N_q}$ instead of at 1 due to the normalization factor multiplied $N_q$ times, yielding the probability of any state being seen being $1/2^{N_q}$.  This does not happen at the larger system sizes as they are limited by shot statistics, but they still trend that direction.  As coupling increases, the curves displaying the Fermi form behavior, except it is now on the $J/h<1$ side of the transition.  As the coupling increases ever larger the curves now seem to collapse more upon one another trending towards an eventual single curve due to there being only one state.  

Comparing across all curve sets for either basis, it is clear that both bases display behavior of the four previously discussed regions in this manuscript for certain parameter ranges.  It is evident that the behavior across distributions with coupling strength is inverted between bases and would be fully shown were each basis' curves taken to an approximate infinite coupling.

\FloatBarrier

\section{Conclusion} 
In this work, we have shown the ability to characterize the shape, behavior, and scaling of cumulative probability distributions to make predictions about necessary shot requirements for large system sizes from distributions of small systems and that this procedure is relevant regardless of the intended model for simulation, despite being reliant on the phase of the model.  The analysis of these probability distributions also yielded an exponential cost that comes with increasing system sizes.  However, this exponential scaling still increases at a fraction of the rate of increase of system size, thus not completely hard capping device usage.  This further motivates the need for intelligent usage of resources, especially as the viability of the quantum simulation of increasingly larger systems becomes more plausible.  We also have shown that this informed shot total prescription can help inform choices for different observables, especially those heavily dependent on entire probability distributions.  In the future, we plan to further quantify the behavior of the cumulative probability distributions, continuing to draw information from them which identifies more about the nature of the simulated systems and looking to further show their broad applicability.  We also look to further understand the ways in which the cumulative distributions, particularly centered around Region II, can be identified with Fermi-Dirac statistical distributions and the implications of that relationship.  

\section{Acknowledgments}
A. K., Z.O., and Y.M. were supported in part by the Dept. of Energy under Award Numbers DE-SC0019139 and DE-SC0026494. We thank the University of Iowa for providing access to the Argon computing facilities.  We thank B. Senseman and M. Asaduzzaman for valuable discussions.  

\section*{Data Availability}  The bitstring dictionaries found from DMRG sampling supporting all analysis and figures are available in the Zenodo repository \cite{zenodo}.

%

\pagebreak

\begin{appendix}

\FloatBarrier
\section{Cumulative Distributions for $R_b/a=3.00$}
\label{app:300}
In this appendix, we briefly highlight results for a Rydberg array with a different $R_b/a$ value, $R_b/a=3.00$.  While not yielding as clean values as $R_b/a=2.00$, the previously performed analysis is still applicable.  

\begin{figure}[!tbp]
    \centering
    \includegraphics[width=8.6cm]{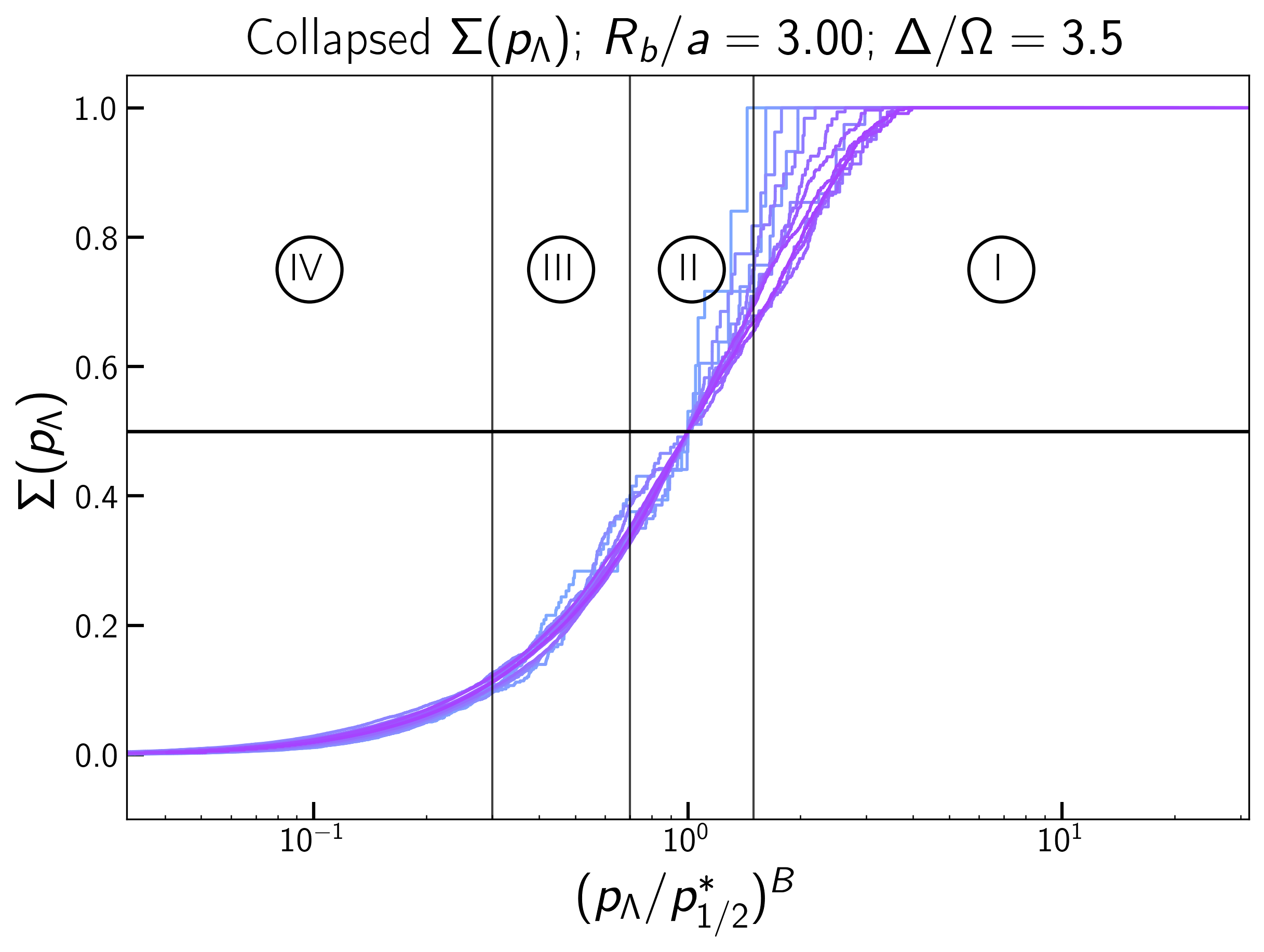}
    \caption{Collapsed cumulative distributions with four regions shown for $R_b/a=3.00$.  }
    \label{fig:linlog_collapse_300}
\end{figure}

As in Figure \ref{fig:linlog_collapse}, the curves are roughly collapsible in Figure \ref{fig:linlog_collapse_300}.  However, it is immediately apparent that the cumulative distribution curves are less uniform across system size, as all these distributions do not fall as neatly upon each other, especially seeing this difference in higher $\Sigma$ values.

\begin{figure}[!tbp]
    \centering
    \includegraphics[width=8.6cm]{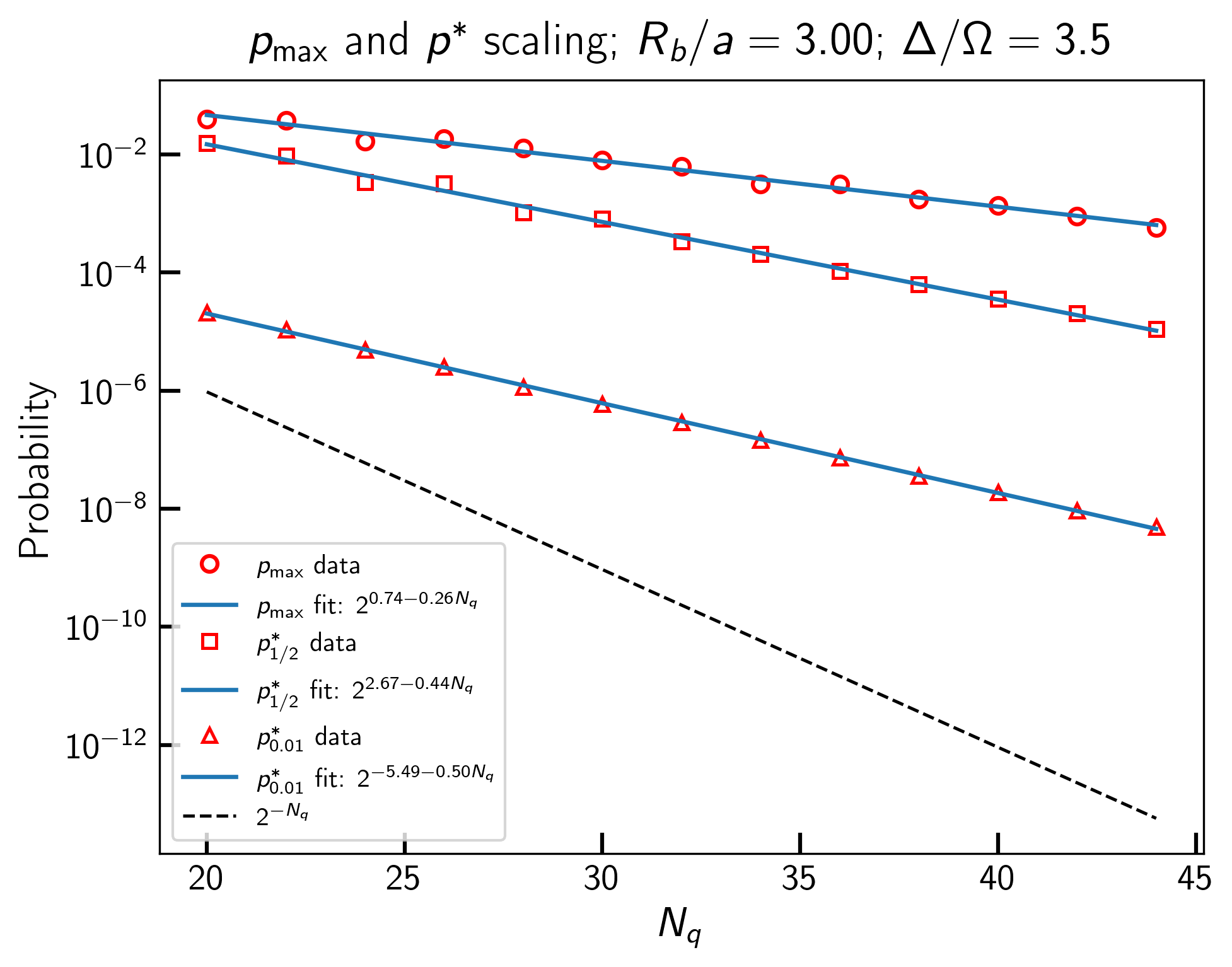}
    \caption{$p_{max}$ and $p^*$ scaling for $R_b/a=3.00$.}
    \label{fig:scaling3}
\end{figure}

\begin{figure}[!tbp]
    \centering
    \includegraphics[width=8.6cm]{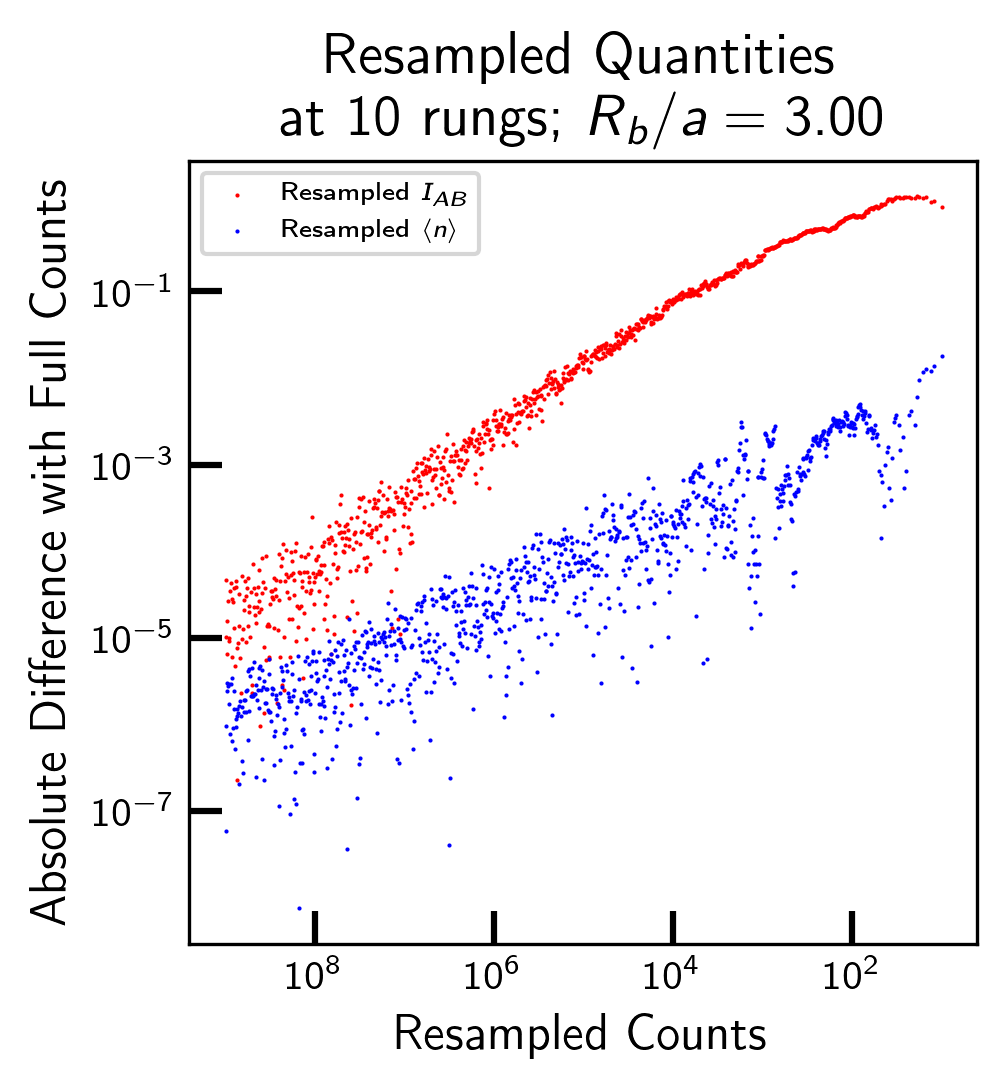}
    \caption{Absolute difference for the observables at the resampled count size and the full counts value ($10^9$) for $R_b/a=3.00$.}
    \label{fig:absdiff_all_300}
\end{figure}

Similarly, Figure \ref{fig:scaling3} shows that all the probabilistic quantities scale similarly for either $R_b/a$ value.  This validates that the methodology we describe is applicable across multiple parameter values for the same model.  This is further verified by looking at the observable plot for $R_b/a=3.00$, Figure \ref{fig:absdiff_all_300}, which also looks very similar to the $R_b/a=2.00$ observable differences plot across resampling.  

\section{Classical Mutual Information as an Approximation for Entanglement}
\label{app:filt}
\begin{figure}[!tbp]
    \centering
    \includegraphics[width=8.6cm]{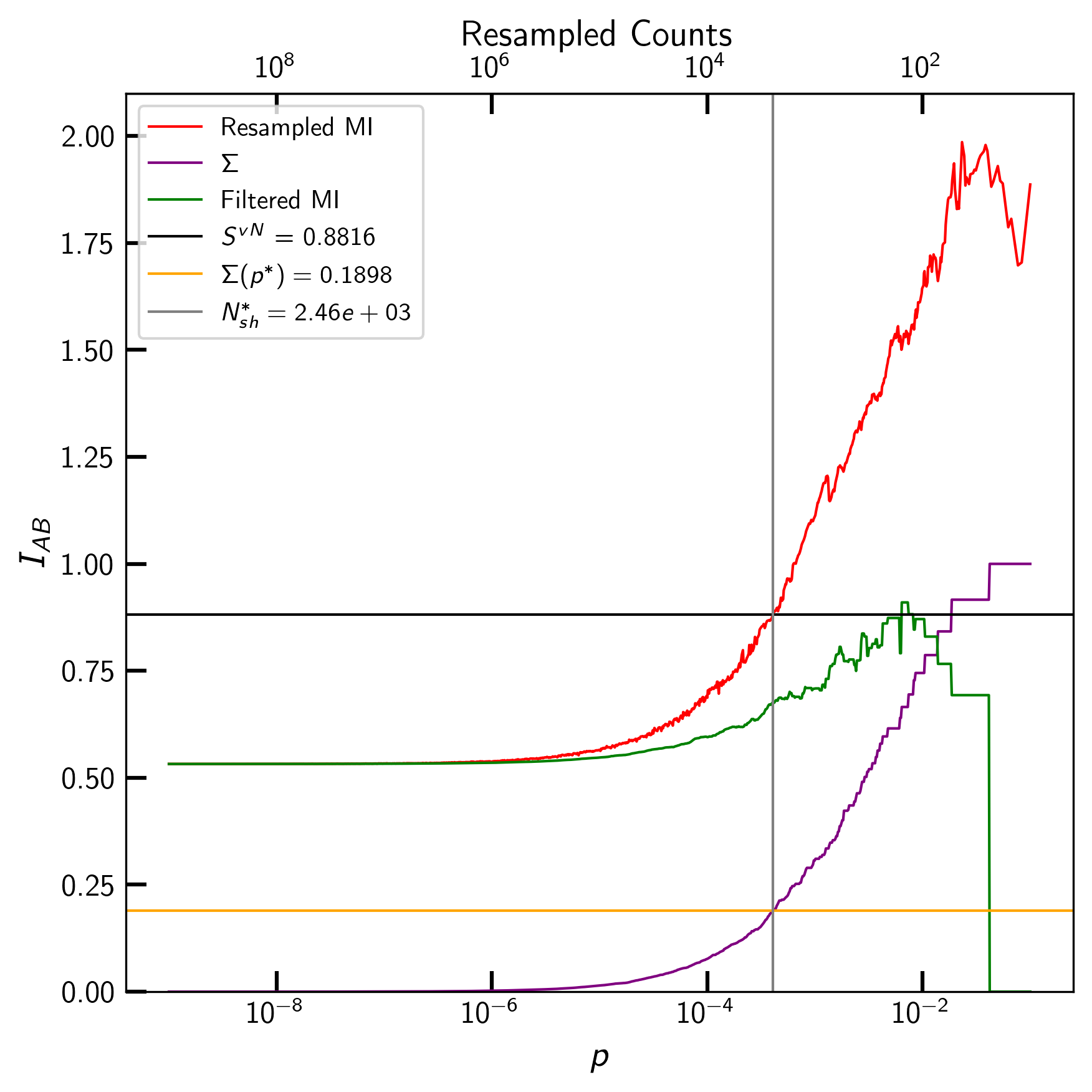}
    \caption{Identifying an optimal number of shots, $N_{sh}^*$, at which the von Neumann entanglement entropy is best approximated for a ten-rung Rydberg atom system with $R_b/a=2.00$ Top x-axis provides resampled counts for the resampled classical mutual information and the bottom x-axis provides probability values for the filtered classical mutual information and cumulative distribution.  }
    \label{fig:fvr}
\end{figure}

Previous works \cite{Kaufman_filter} have shown that by removing low probability states below a certain threshold in a probability distribution, a process referred to as filtering, the classical mutual information can better approximate the von Neumann entanglement entropy than when calculated with a full distribution where it is just a strict lower bound.  Taking inspiration from this previous study, motivated by the goal of having experiments informed by cumulative probability distributions, we look to see if there could be an optimal number of shots that best approximate the von Neumann entanglement.  This would be distinctly different from the procedure outlined in \cite{Kaufman_filter}, which systematically removed lower probabilities from the distribution and renormalizing then calculating the classical mutual information.  What we look to find is if there is a probability distribution at some shot total that best approximates the classical mutual information, where no added procedure is performed on the probability distribution, which retains the respective scaling of probability values as the exact state.  This attempt is shown in Figure \ref{fig:fvr} for a ten-rung Rydberg atom system at $R_b/a=2.00$. 

The resampled mutual information rapidly rises above its original full sample ($10^9$ counts) value.  This mirrors the filtered mutual information, despite increasing to higher values.  Taking advantage of already knowing the entanglement, we identify where it intersects the resampled mutual information.  We then identify the resampled count value where this occurs, $N_{sh}^*$, and also identify what the cumulative distribution value of intersection is, $\Sigma(p^*)$.  For ten rungs, in this model and parameter regime, we find that the entanglement entropy could be best approximated where $\Sigma$ is only at about a fifth of its maximum and only necessitating a few thousand, $N_{sh}^*\approx2400$, shots.  These results are encouraging, indicating that given fairly accurate experiments, using this method the entanglement entropy could be approximated with relatively modest resources.  

\pagebreak

\section{Dependence of Parameter Fitting}
\label{app:slope}
\begin{figure*}[t]
    \centering
    \includegraphics[width=.48\linewidth]{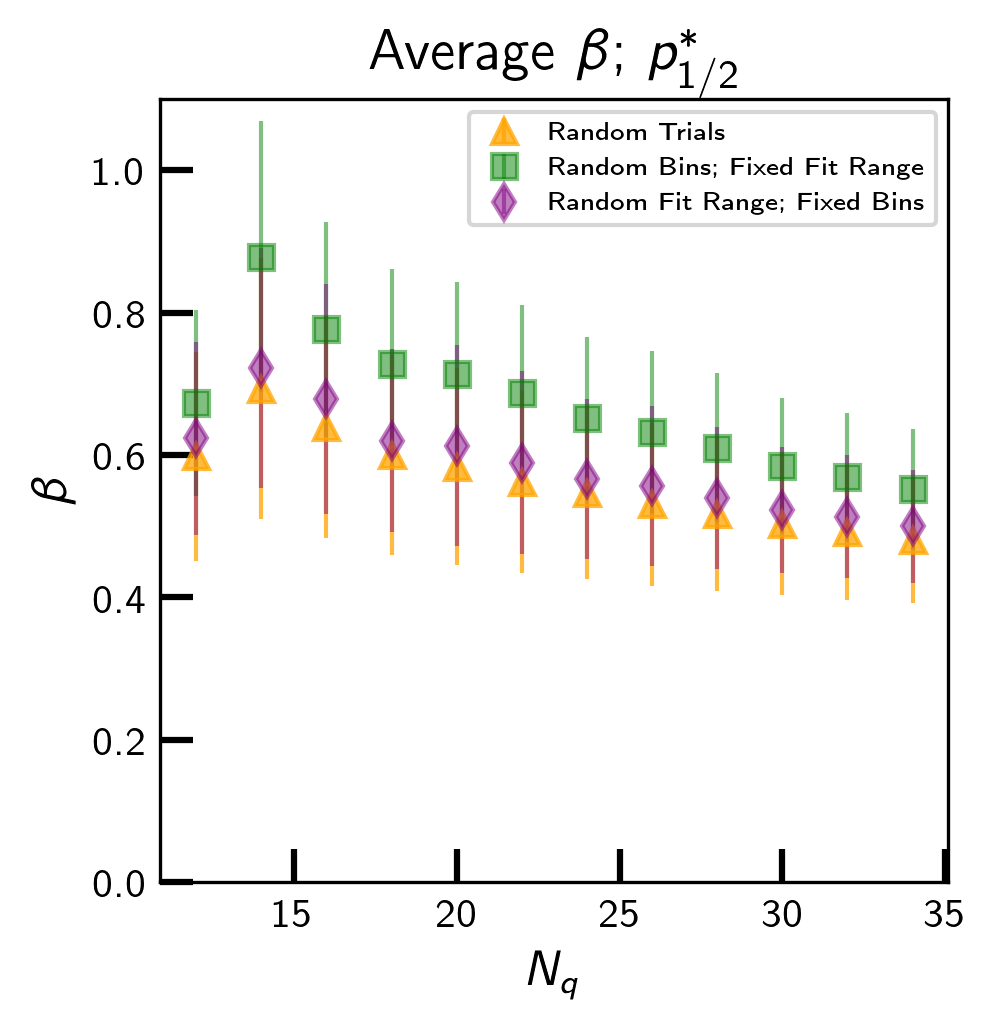}
    \includegraphics[width=.48\linewidth]{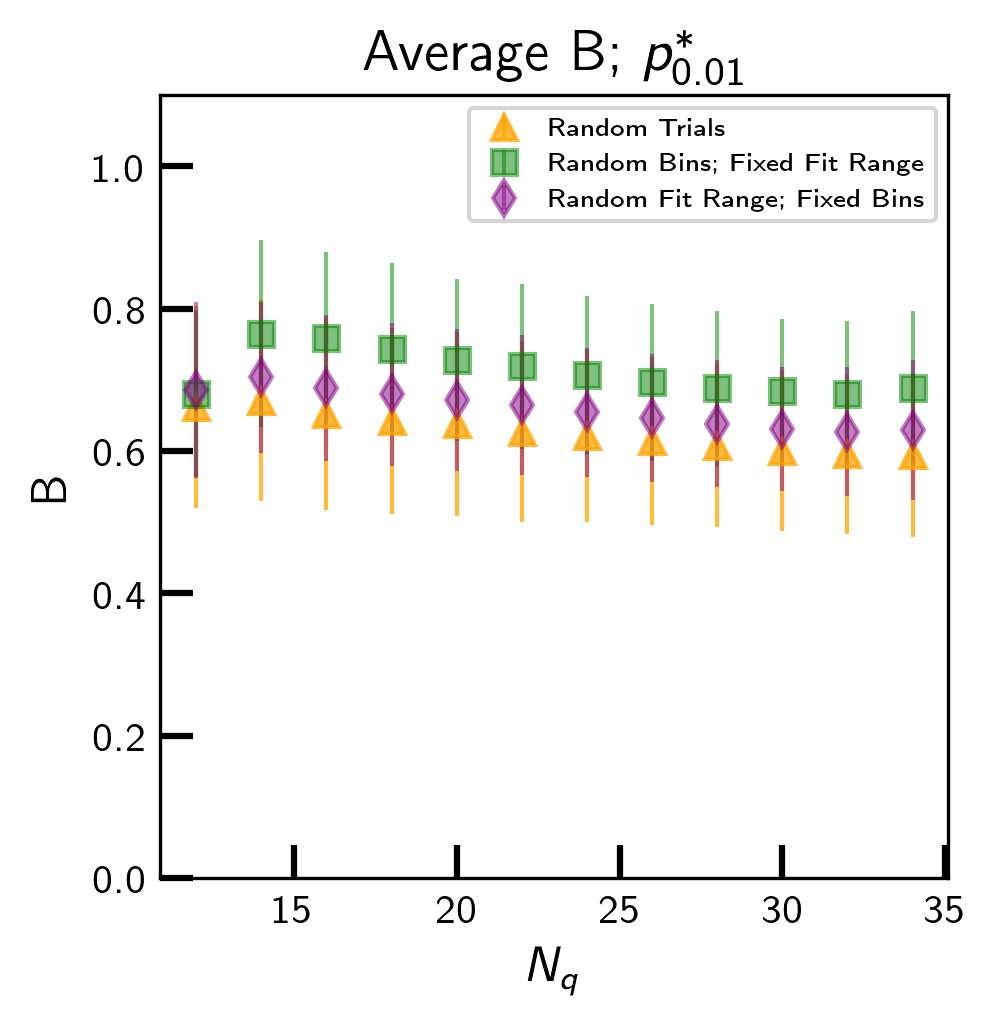}
    \caption{On the left: statistics on slope fitting in Region II.  On the right: statistics on slope fitting in Region III.  Points are averages and error bars are standard deviation across trials.  }
    \label{fig:slope_stats}
\end{figure*}

When calculating the slopes for the fits in Regions II and III, there are computational choices made which can contribute to the value of the slope.  First, the probabilities used in the cumulative distributions come from matrix product state sampling. Therefore, there are thousands of different bitstring values and companion probabilities that contribute to the distributions.  When finding the distributions, a set of $p_\Lambda$ are chosen to use as the precise steps to calculate $\Sigma$ at, this is the binning of the cumulative distributions.  Second, the range of points over which to calculate the fit is an independent choice.  Both the binning and fit range of points around $p^*$ will directly impact the $\beta$ or $B$ values in Regions II and III respectively.  

To quantify how both the binning and the fit ranges can impact the calculation of these values, we have calculated the fits over a variety of system sizes while fixing either the binning or fit range then varying over the other.  With this method, we are able to calculate average values of the parameters and standard deviations on those values.  Furthermore, to find a range over which these slope values vary, we complete a third procedure where we randomly select a number of bins and a range of fitting within that binning.  The results of these procedures are shown for $\beta$ with $p^*_{1/2}$ in the left plot of Figure \ref{fig:slope_stats} and for $B$ with $p^*_{0.01}$ in the right plot.

These plots reveal that there is a higher sensitivity to manipulating binning than overall fitting range.  By allowing for more bins, the cumulative distribution curves can be exact.  Meaning, any steps in the curves are due to actual discrete steps in the probability distribution as opposed to forcing nearby probabilities into the same bins.  Further, less of a sensitivity to fitting range indicates that the curves are necessarily smooth in Regions II and III such that even smaller fitting ranges can provide consistent values. 
To avoid any limitations in binning, perform what is referred to as exact binning: where instead of enforcing a number of bins to fit all probabilities into, rather each individual bitstring provides a discrete step in the cumulative distribution.  Broadly speaking, for both the $\beta$ and $B$ fits, there is less variance in slope values as system 
size increases.  This follows from the apparent smoothness of the distribution curves as system size increases as shown in the main text.  Additionally, it seems the $B$ values of the right plot in Figure \ref{fig:slope_stats} are more robust against all procedures than the more widely varying $\beta$ values of the left plot.

\end{appendix}

\end{document}